\documentclass[floatfix, noshowpacs, preprintnumbers, twocolumn, amsmath, amssymb, aps, prx, superscriptaddress]{revtex4-1}

\usepackage{graphicx, xcolor}
\usepackage{soul}
\usepackage{times}
\usepackage{multirow}
\usepackage{booktabs}
\usepackage{hhline}
\usepackage{rotating}
\usepackage{complexity}
\usepackage{color}

\begin{document}

\title{Pattern recognition techniques for Boson Sampling validation}

\author{Iris Agresti}
\affiliation{Dipartimento di Fisica, Sapienza Universit\`{a} di Roma,
Piazzale Aldo Moro 5, I-00185 Roma, Italy}

\author{Niko Viggianiello}
\affiliation{Dipartimento di Fisica, Sapienza Universit\`{a} di Roma,
Piazzale Aldo Moro 5, I-00185 Roma, Italy}

\author{Fulvio Flamini}
\affiliation{Dipartimento di Fisica, Sapienza Universit\`{a} di Roma,
Piazzale Aldo Moro 5, I-00185 Roma, Italy}

\author{Nicol\`o Spagnolo}
\affiliation{Dipartimento di Fisica, Sapienza Universit\`{a} di Roma,
Piazzale Aldo Moro 5, I-00185 Roma, Italy}

\author{Andrea Crespi}
\affiliation{Istituto di Fotonica e Nanotecnologie, Consiglio
Nazionale delle Ricerche (IFN-CNR), Piazza Leonardo da Vinci, 32,
I-20133 Milano, Italy}
\affiliation{Dipartimento di Fisica, Politecnico di Milano, Piazza
Leonardo da Vinci, 32, I-20133 Milano, Italy}

\author{Roberto Osellame}
\affiliation{Istituto di Fotonica e Nanotecnologie, Consiglio
Nazionale delle Ricerche (IFN-CNR), Piazza Leonardo da Vinci, 32,
I-20133 Milano, Italy}
\affiliation{Dipartimento di Fisica, Politecnico di Milano, Piazza
Leonardo da Vinci, 32, I-20133 Milano, Italy}

\author{Nathan Wiebe}
\affiliation{Station Q Quantum Architectures and Computation Group, Microsoft Research, Redmond, WA, United States}

\author{Fabio Sciarrino}
\affiliation{Dipartimento di Fisica, Sapienza Universit\`{a} di Roma,
Piazzale Aldo Moro 5, I-00185 Roma, Italy}

\begin{abstract}
The difficulty of validating large-scale quantum devices, such as Boson Samplers, poses a major challenge for any research program that aims to show quantum advantages over classical hardware. Towards this aim, we propose a novel data-driven approach wherein models are trained to identify common pathologies using unsupervised machine learning methods.  We illustrate this idea by training a classifier that exploits $K$-means clustering to distinguish between Boson Samplers that use indistinguishable photons from those that do not.  We tune the model on numerical simulations of small-scale Boson Samplers and then validate the pattern recognition technique on larger numerical simulations as well as on photonic chips in both traditional Boson Sampling and scattershot experiments. The effectiveness of such method relies on particle-type-dependent internal correlations present in the output distributions. This approach performs substantially better on the test data than previous methods and underscores the ability to further generalize its operation beyond the scope of the examples that it was trained on.
\end{abstract}

\maketitle

\section{Introduction}
There has been a flurry of interest in quantum science and technology in recent years that has been focused on the transformative potential that quantum computers have for cryptographic tasks~\cite{shor_97}, machine learning \cite{MariaShuld, Biamonte} and quantum simulation \cite{Feynman, UQS}.  While existing quantum computers fall short of challenging their classical brethren for these tasks, a different goal has emerged that existing quantum devices could address: namely, testing the extended Church-Turing thesis.  The extended Church-Turing thesis is a widely held belief that asserts that every physically reasonable model of computing can be efficiently simulated using a probabilistic Turing machine.  This statement is, of course, controversial since, if it were true, then quantum computing would never be able to provide exponential advantages over classical computing.  Consequently, providing evidence that the extended Church-Turing thesis is wrong is more philosophically important than the ultimate goal of building a quantum computer.

\begin{figure}[b!]
\centering
\includegraphics[width=0.49\textwidth]{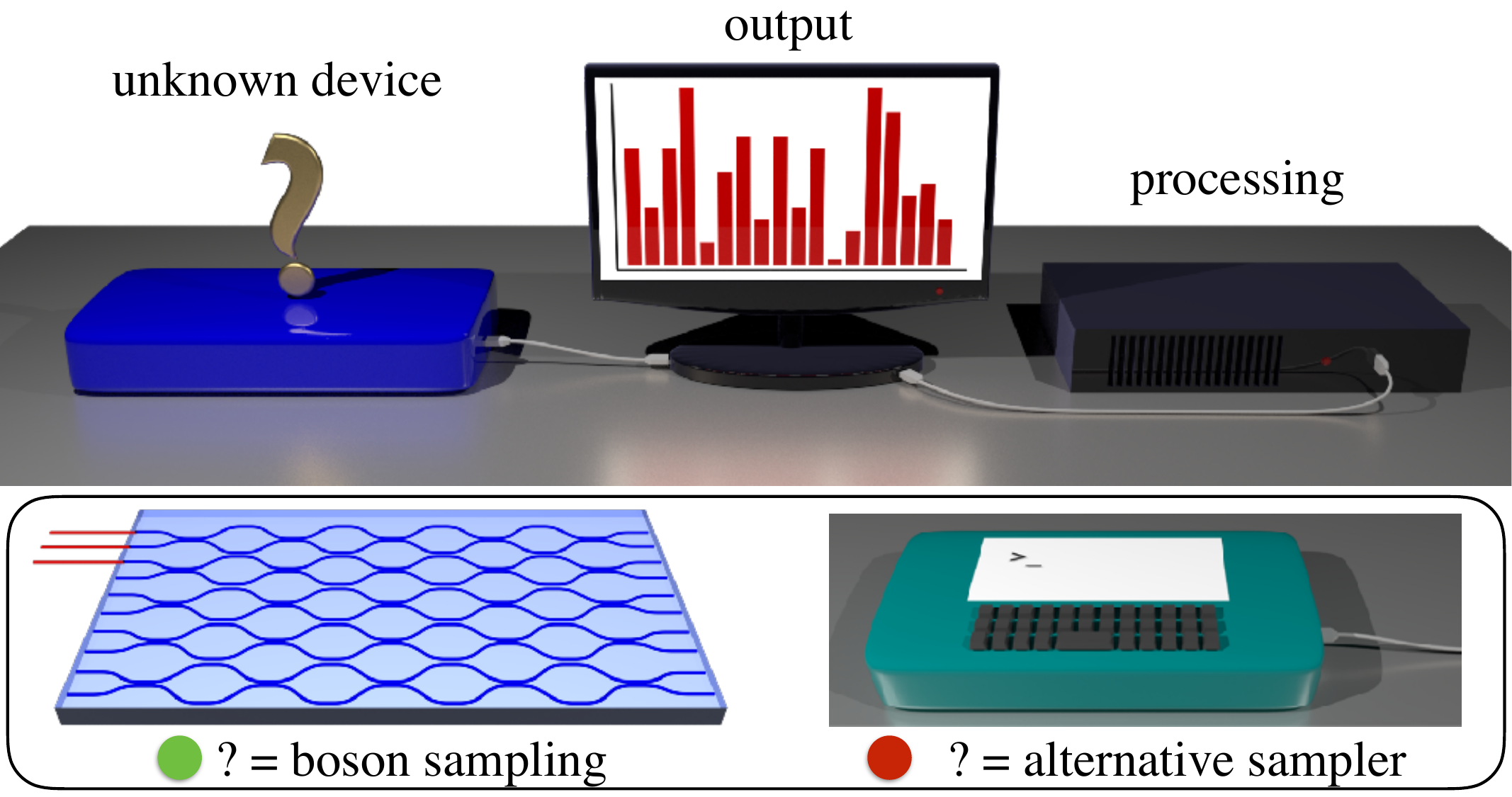}
\caption{{\bf Validation of Boson Sampling experiments.} An agent has to discriminate whether a finite sample obtained from an unknown device has been generated by a quantum device implementing the Boson Sampling problem or by an alternative sampler.}
\label{fig:figure1}
\end{figure}

Various intermediate computing models have been proposed in the last few years, that promise to be able to provide evidence of a quantum computational supremacy, namely the regime where a quantum device starts outperforming its classical counterpart in a specific task. Such models mostly belong to the category of sampling problems, i.e. simulating the distribution sampled from a quantum system, that is believed to be classically hard to compute. These include quantum circuits with commuting gates \cite{Bremner16, Gao17, Bremner16arxiv}, quantum simulators with fully certifiable final states \cite{Bermejo-Vega17} and quantum random circuits \cite{Boixo2016}.

A significant step in this direction has been achieved in particular by Aaronson and Arkhipov \cite{AA10} with the formal definition of a dedicated task known as Boson Sampling. This is a computational problem that consists in sampling from the output distribution of $n$ indistinguishable bosons evolved through a linear unitary transformation. This problem has been shown to be classically intractable (even approximately) under mild complexity theoretic assumptions. Indeed, the existence of a classical efficient algorithm to perform Boson Sampling would imply the collapse of the polynomial hierarchy to the third level \cite{AA10}. Such a collapse is viewed among many computer scientists as being akin to violating the laws of thermodynamics. Thus demonstrating that a quantum device can efficiently perform Boson Sampling is powerful evidence against the extended Church-Turing thesis.  Furthermore, the simplicity of Boson Sampling has already allowed experiments at a small scale with different photonic platforms \cite{Broome2013,Spring2013,Till2012,Crespi2012,Spagnolo13a,Spagnolo2013a,Carolan2013a,lund2014, scattershot2015,Carolan2015,Loredo2016,He2016,Wang2016,Wang2018} and also alternative approaches have been proposed, for example exploiting trapped ions \cite{trapped_ions} or applying random gates in superconducting qubits \cite{Boixo2016}.

Despite the fact that Boson Sampling is within our reach, a major caveat remains.  The measurement statistics for Boson Samplers are intrinsically exponentially hard to predict.  This implies that, even if someone manages to build a Boson Sampler that operates in a regime beyond the reach of classical computers, then the experimenter needs to provide evidence that their Boson Sampler functions properly for the argument against the extended Church-Turing thesis to be convincing.  This task is not straightforward in general for large quantum systems~\cite{Gogolin2013,aaronson2013bosonsampling,wiebe2014using,Aolita2015} and it represents a critical point for all the above-mentioned platforms seeking a first demonstration of quantum supremacy. 
Indeed, to conclusively certificate a sampler using a conventional metric, such as a cross-entropy, it is necessary to estimate the probability that the true Boson sampler yields the outcome probability. However, computing such a probability efficiently would imply BQP = \#P, which is totally implausible from a complexity-theoretic standpoint even given a quantum computer. Hence, one could consider providing progressively more stringent tests able to exclude relevant alternative error models.
A first approach to ensure quantum interference could involve testing pairwise mutual indistinguishability by two-photon Hong-Ou-Mandel experiments \cite{hom1987}, however such method fails to completely characterize multiphoton interference \cite{mens2017}. While techniques exist that use likelihood ratios to validate~\cite{Spagnolo2013a,Bentivegna2014, Viggianiello17tvd}, they work only for small systems.  Other existing techniques exploit statistical properties of bosonic states~\cite{Carolan2013a,Shchesnovich2016,Liu2016,Walschaers2016,Benti2016,Giordani2018} or symmetries of certain Boson Samplers~\cite{Tichy2013,Crespi2015,Carolan2015,Crespi2016,Dittel17,Dittel18,Viggianiello17sys}, however these methods are much more limited in scope.

In this article we devise a prototypical methodology based on machine learning (ML) techniques to detect known types of malfunctions occurring in a quantum hardware performing sampling tasks. We apply the test here in the context of Boson Sampling, however the intuition can be reformulated for other problems. This method compares features of the collected data sample with those of a second one obtained from a reference distribution, evaluating their compatibility. By generating the reference sample from an efficiently computable distribution corresponding to a well-defined pathology of the problem, it is then possible to exclude that such pathology is observed in the measured sample. More specifically, building on results of Wang and Duan \cite{Wang_Duan}, we devise a compatibility test between a trusted Boson Sampler and an untrusted device that looks at data structure in a suitable space (see Fig. \ref{fig:figure1}). We then test experimentally our method both on traditional Boson Sampling and scattershot Boson Sampling \cite{lund2014, scattershot2015}, showing that the algorithm is able to identify relevant pathologies in the measured samples. Finally, we provide a physical insight on the mechanism behind the functioning of the proposed clustering-based method in the investigated Boson Sampling framework. Thanks to their versatility and their capability to operate without an in-depth knowledge of the physical system under investigation, clustering techniques may prove effective even in a scope broader than Boson Sampling \cite{Bremner16, Boixo2016, Gao17, Bremner16arxiv, Bermejo-Vega17}.

\section{Boson Sampling and its validation}
Before going into detail about our approach, we need to discuss the Boson Sampling problem at a more technical level.  Boson Sampling is a computational problem \cite{AA10} that corresponds to sampling from the output probability distribution obtained after the evolution of $n$ identical, i.e. indistinguishable, bosons through a $m$-mode linear transformation. Inputs of the problem are a given $m \times m$ Haar-random unitary matrix $U$, describing the action of the network on the bosonic operators according to the input/output relation $a_{i}^{\dag} = \sum_{j} U_{i,j} b_{j}^{\dag}$, and a mode occupation list $S=\{s_{1},\ldots,s_{m}\}$ where $s_{i}$ is the number of bosons on input mode $i$, being $\sum_{i} s_{i} = n$. For $m\gg n^{2}$ and considering the case where at most one photon is present in each input ($s_{i} = \{0,1\}$) (collision-free scenario), sampling, even approximately, from the output distribution of this problem is classically hard. Indeed, in this regime for $(n,m)$ the probability of a collision event becomes negligible and thus the only relevant subspace is the collision-free one \cite{AA10,Spagnolo13a}. The complexity, in $n$, of the known classical approaches to perform Boson Sampling relies on the relationship between input/output transition amplitudes and therefore on the calculation of permanents of complex matrices, which is $\#$P-hard \cite{valiant}. More specifically, given an input configuration $S$ and an output configuration $T$, the transition amplitude $\mathcal{A}_{U}(S,T)$ between these two states is obtained as $\mathcal{A}_{U}(S,T)=\mathrm{per}(U_{S,T})/(s_{1}!\ldots s_{n}!\hspace{1mm} t_{1}! \ldots t_{n}!)^{1/2}$, where $\mathrm{per}(U_{S,T})$ is the permanent of the $n \times n$ matrix $U_{S,T}$ obtained by selecting columns and rows of $U$ according to the occupation lists $S$ and $T$ \cite{scheel}. 

From a practical perspective, an essential aspect of any algorithm is that of verifying the correctness of its outputs.  Verification can be trivial, as for factoring \cite{shor_97}, or exponentially hard, as for Boson Sampling. In the latter case, which likely involves the evaluation of permanents, this stage is commonly referred to as \emph{certification}. A similar goal, which aims to reduce the required physical resources in view of large-scale applications, is instead that of \emph{validation}. In this case, one does not exactly attempt to verify the correctness of an outcome but, rather, to exclude known undesired models for the system that produced it. This approach has the advantage of being able to rule out many of the most likely pathologies that a Boson sampler can face, without impacting the algorithm's practicality.

Thus, due to the complexity of evaluating the permanent, it is necessary to identify methods that do not require the calculation of input/output probabilities to validate the functioning of the device. Furthermore, in a quantum supremacy regime the number of input/output combinations becomes very large, since it scales as $\binom{m}{n}$. Hence it is also necessary to develop suitable techniques that are tailored to deal with a large amount of data. In Tab. \ref{tableValidation} we report a summary of the currently developed techniques, highlighting their main features and performances, in comparison with the method proposed in this article.

\begin{table*}[ht!]
\begin{center}
\includegraphics[width=0.8\textwidth]{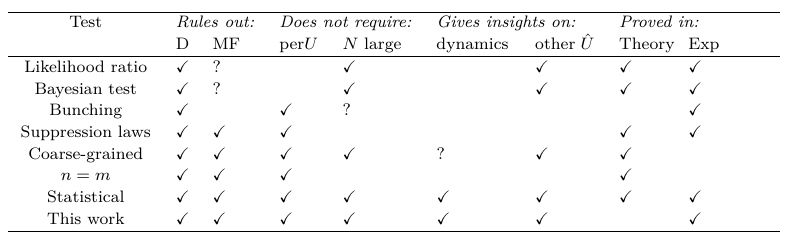}
\end{center}
\caption{Summary of the available protocols to validate quantum interference (Q) against experiments with distinguishable photons (D) or mean-field states (MF). Ideal validations should be reliable and efficient, meaning that they should not require resources exponential in the size of the problem, neither computational (e.g. the evaluation of permanents) nor physical  (e.g. in the number of samples $N$). Also, they could provide insights on the multi-photon dynamics for a given transformation, as well as being applicable to conditions other than Q, D or MF.
LR -- Likelihood ratio \cite{Spagnolo2013a, Wang2016, He2016, Wang2018, scattershot2015}; ZTL -- zero-transmission law (or suppression law) \cite{Tichy2013, Dittel17, Dittel18, Crespi2015, Crespi2016, Carolan2015, Viggianiello17sys, Viggianiello17tvd}; Bayesian \cite{Bentivegna2014, Wang2016, He2016, Wang2018, Viggianiello17tvd}; Bunching \cite{Carolan2013a, Shchesnovich2016}; CG -- Coarse-graining \cite{Wang_Duan}; n=m  \cite{Liu2016}; Statistical Benchmark  \cite{Walschaers2016, Giordani2018}.}
	\label{tableValidation}
\end{table*}

\begin{figure}[t!]
	\includegraphics[width=0.5\textwidth]{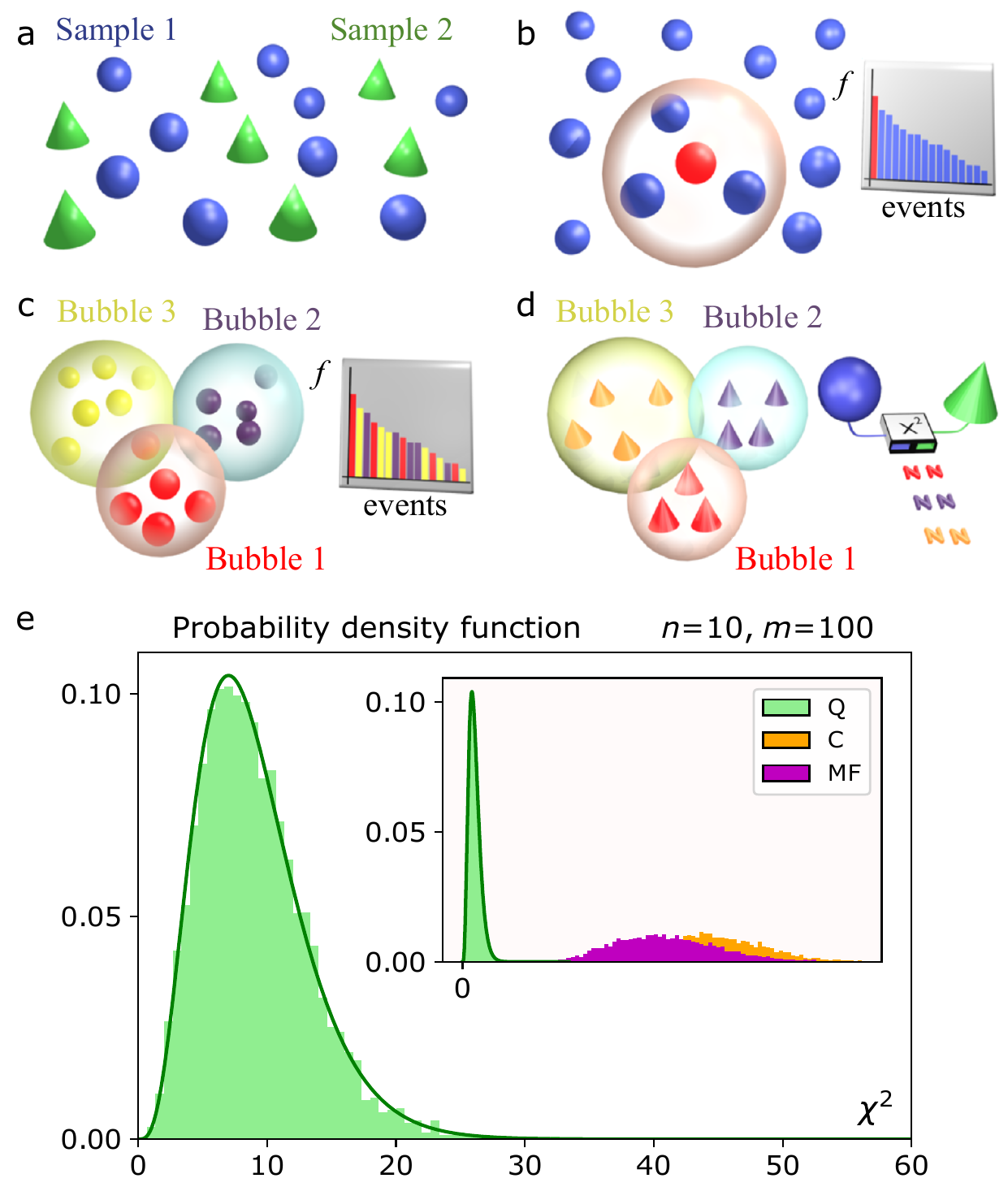}
	\caption{\textbf{Pattern recognition techniques for validation} \textbf{a}, A sample is drawn from each of the two Boson Samplers to be compared. The events belonging to one of the two samples are partitioned according to the criteria of the pattern recognition technique. \textbf{b-c}, Bubble clustering sorts the events according to their observation frequency and the state with the highest frequency is chosen as the center of the first cluster: all events with distance from the center smaller than a cut-off radius $\rho_{1}$ are included in this cluster (b). Then, starting from the unassigned events, this procedure is iterated until all of the observed events are included in some bubble. At this point, each cluster is characterized by a center and a radius (c). \textbf{d}, The observed events belonging to the second sample are classified by using the structure tailored from the first sample: each event belongs to the cluster with the nearest center. {\bf e}, A $\chi^2$ test with $\nu=N_\textup{bubbles}-1$ degrees of freedom is performed (here using \textit{K-means}) to compare the number of events belonging toeach of the two samples (green: quantum -Q- with indistinguishable photons; orange: classical -C- with distinguishable photons; purple: Mean-Field state) by using the obtained cluster structure. This variable quantifies the compatibility between the samples.}
	\label{bubble:clustering}
\end{figure}

\section{Pattern recognition techniques for validation}
In the regime where a Boson Sampling device is expected to outperform its classical counterpart, the validation problem has inherently to deal with the exponential growth of the number of input/output combinations. A promising tool in this context is provided by the field of machine learning, which studies how a computer can acquire information from input data and learn to make data-driven predictions or decisions \cite{Simon2013}. Significant progresses have been achieved in this area over the past few years \cite{Bishop2006,murphy2012}. One of its main branches is represented by unsupervised machine learning, where dedicated algorithms have to find an inner structure in an unknown data set. One of the main unsupervised learning approaches is clustering, where data are grouped in different classes according to collective properties recognized by the algorithm \cite{maimon05}. Since this approach is designed to identify hidden patterns in a large amount of data, clustering techniques are promising candidates to be applied for the Boson Sampling validation problem.

Let us discuss the general scheme of the proposed validation method based on pattern recognition techniques. This approach allows us to employ various clustering methods within the protocol, which allow us to choose the method that optimizes the performance on the training data. Given two samples obtained respectively from a \textit{bona fide} Boson Sampler, that is a trusted device, and a Boson Sampler to be validated, the sequence of operations consists in (i) finding a cluster structure inside the data belonging to the first sample, (ii) once the structure is completed, organizing the data of the second sample by following the same structure of the previous set, and (iii) performing a $\chi^{2}$ test on the number of events per cluster for the two independent samples. The $\chi^{2}$ variable is evaluated as $\chi^2=\sum_{i=1}^{N_c}\sum_{j=1}^2\frac{(N_{ij}-E_{ij})^2}{E_{ij}}$, where index $j$ refers to the samples, index $i$ to the $N_c$ clusters, $N_{ij}$ is the number of events in the $i$-th cluster belonging to the $j$-th sample and $E_{ij}$ is the expected value of observed events belonging to the $j$-th sample in the $i$-th cluster $E_{ij}= N_{i}N_{j}/{N_c}$, with $N_{i}=\sum_{j=1}^2 N_{ij}$, $N_{j}=\sum_{i=1}^{N_c} N_{ij}$ and $N=\sum_{i=1}^{N_c} \sum_{j=1}^2 N_{ij}$. If the null hypothesis of the two samples being drawn from the same probability distribution is correct, the evaluated variable must follow a $\chi^2$-distribution with $\nu$ = $N_\mathrm{c}-1$ degrees of freedom. This scheme can be applied by adopting different metric spaces and different clustering techniques. Concerning the choice of the metric, both 1-norm and 2-norm distances can be employed as distance $d$ between two Fock states $\Psi$ and $\Phi$, namely $d=L_1=\sum_{i=1}^M \vert \Psi_i-\Phi_i \vert$ or $d=L_{2}= \sqrt{\sum_{i=1}^{M}|\psi_{i}-\phi_{i}|^2}$, with $\Psi_i$ and $\Phi_i$ being respectively the occupation numbers of $\Psi$ and $\Phi$ in the $i$-th mode. 

\section{Adopted clustering techniques} 

Several clustering methods were employed within our validation scheme: (a) a recent proposal by Wang and Duan \cite{Wang_Duan}, whose concept is shown in Fig. \ref{bubble:clustering}, and two unsupervised machine learning techniques, (b) \textit{agglomerative Hierarchical Clustering} and (c) \textit{K-means clustering}. Two variations of the latter approach were also examined, to increase the strength of our model. A short description of each adopted method follows briefly.

(a) The protocol proposed by Wang and Duan \cite{Wang_Duan}, and hereafter named \textit{bubble clustering}, determines the inner cluster structure of a sample by (i) sorting in decreasing order the output events according to their frequencies, (ii) choosing the observed state with the highest frequency as the center of the first cluster, (iii) assigning to such cluster all the states belonging to the sample whose distance $d$ from its center is smaller than a cutoff radius $\rho_i$, and (iv) iterating the procedure with the remaining states until all the observed events are assigned.

(b) \textit{Hierarchical clustering}, in its bottom-up version, starts by assigning each observed event to a separate class. Then, the two nearest ones are merged to form a single cluster. This grouping step is iterated, progressively reducing the number of classes. The agglomeration stops when the system reaches a given halting condition pre-determined by the user. In the present case, the algorithm halts when no more than 1$\%$ of the observed events is included in some cluster containing less than 5 events (See Supplemental Material). All of these smallest clusters are considered as outliers and removed from the structure when performing the $\chi^{2}$ test. The distance between two clusters is evaluated as the distance between their centroids. The centroid of a cluster is defined as the point that minimizes the mean distance from all the elements belonging to it.

(c) \textit{K-means} is a partitioning clustering algorithm where the user has to determine the number of classes ($k$) \cite{mcq:km, forgy:km,lloyd:km}. With this method, the starting points for centroid coordinates are chosen randomly. Then, two operations are iterated to obtain the final cluster structure, that are selecting elements and moving centroids. The first one consists in assigning each observed event to the cluster whose centroid has the smallest distance from it. Then, once the $k$ clusters are completed, the centroid of each cluster is moved from the previous position to an updated one, given by the mean of the elements coordinates. These two operations are repeated until the structure is stable. Given a set of $k$ centroids ($c_{1},...c_{k}$) made of ($N_{1},.. N_{k}$) elements ($e_{11},... e_{1n_{1}}, ..., e_{k1},..., e_{k n_{k}}$), where $\sum_{i=1}^{k}N_{i}$= $n$, the operations of selecting elements and moving the centroids minimize the objective function $\frac{1}{N_c} \sum_{j=1}^{k}\sum_{i=1}^{N_{k}} d(e_{ij},c_{j})$. Several trials were made to determine the optimal number of clusters, showing that the performance of the test improves for higher values of $k$ and then reaches a constant level. We then chose to balance the two needs of clusters made up of at least 5 elements, since the compatibility test requires a $\chi^{2}$ evaluation, and of a high efficacy of the validation test (See Supplemental Material). 

\section{Variations of K-means clustering}

With \textit{K-means} different initial conditions can lead to a different final structure. Hence, the algorithm can end up in a local minimum of its objective function. To avoid this issue, we considered three different strategies: (I-II) replacing the random starting condition with two initialization algorithms, namely (I) \textit{K-means++} and (II) a preliminary run of \textit{Hierarchical clustering}, and (III) building on the same data set several cluster structures.
(I) Once the user has set the number of clusters $k$, the first center is picked uniformly among the observed events. Then, for each observed element $e$ the distance $d(e)$ from the nearest of the picked centroids is evaluated. A new centroid is subsequently chosen randomly among the observed events, by assigning to each one a different weight given by $d(\textit{e})^2$. This procedure is iterated until all $k$ cluster centers are inizialized. Then, standard \textit{K-means clustering} can be applied.
(II) The user has to set the halting condition for \textit{Hierarchical clustering}. As discussed previously, in our case the process is interrupted when the fraction of outliers is smaller than a chosen threshold condition ($\le$ 0.01). The centroids of the final cluster structure obtained from \textit{Hierarchical clustering} are used as starting condition for \textit{K-means}.
(III) As said, when adopting \textit{K-means clustering} the final structure is not deterministic for a given data set. Hence, to reduce the variability of the final condition and thus avoid the algorithm to get stuck in a local minimum, the \textit{K-means} method is run an odd number of times (for instance 11) and majority voting is performed over the compatibility test results.
Finally, the adoption of \textit{K-means++} (I) and majority voting (III) can also be simultaneously combined.

\section{Benchmarking the protocol}
\label{numerical_results}

As a first step, we performed a detailed analysis to identify the most suitable among the mentioned clustering algorithms. More specifically, we proceeded with the two following steps: (i) a \textit{tuning stage} and (ii) a \textit{cross-validation stage}. The figure of merit quantifying the capability of each test to perform correct decisions is the success percentage, i.e., the probability that two samples drawn from the same statistical population are labeled as compatible while two samples drawn from different probability distributions are recognized as incompatible.

\begin{table}[t!]
	\centering
    \footnotesize
	{\renewcommand\arraystretch{1.45}
		\begin{tabular}{|ccc|c|c||c|c||c|c}
			\multicolumn{3}{c}{}                                                                                                                   & \multicolumn{4}{c}{\textbf{Output Classification}}                                                                          &  \multicolumn{1}{c}{}     &  \\ 	\cline{4-7}
			\multicolumn{3}{c}{}                                                                                                                   & \multicolumn{2}{|c||}{\textit{\textbf{1-norm}}} 	 & \multicolumn{2}{c|}{\textit{\textbf{2-norm}}}	 &  \multicolumn{1}{c}{}            &   \\  \cline{4-7} 
			\multicolumn{2}{c}{}   &                                                                                                               &   \textit{Ind.}   &  \textit{Dis.}     & \multicolumn{1}{c|}{\textit{Ind.}}    & \multicolumn{1}{c|}{\textit{Dis.}}    & \multicolumn{1}{c}{} & \\                
			\hhline{--------}
			&\textbf{(a) Bubble}   &   &  95  &  5 &   96  & 4  &\textit{Ind.}   &      \\        
			\cline{4-8}
		& \textbf{clustering}  &   &  33  &  67 & 31 & 69  &   \textit{Dis.} &   \\ 
			\hhline{========}
		& \textbf{(b) Hierarchical}   &   & 1  & 99  &  8  & 92   &  \textit{Ind.}    &   \\  
			\cline{4-8}
		     & \textbf{clustering}  &  & 2 &  98  &  5 & 95  & \textit{Dis.}   &            \\ 
			\hhline{========}
		\multirow{5}{12pt}{} &
			\multicolumn{1}{|c|}{\multirow{3}{*}{\textbf{Uniformly distributed}}} & \multirow{2}{*}{\textit{Single trial}}    &  98  &  2   &  95  &  5   &    \textit{Ind.}  &  \\  \cline{4-8}
			 \multicolumn{1}{|c}{\textit{}} & \multicolumn{1}{|c|}{\textit{}} & &  10  & 90  &  21  & 79  &  \textit{Dis.}   &    \\ 
			\hhline{|~|~|------|}
			\textit{} & \multicolumn{1}{|c|}{\multirow{1}{*}{\textbf{initialized centroids}}} & \multirow{1}{*}{\textit{Majority}}    & 100    & 0  &   99   & 1  & \textit{Ind.} &  \\ 
			\cline{4-8}
			 \multicolumn{1}{|c|}{} & & \multicolumn{1}{|c|}{\multirow{1}{*}{\textit{voting}}}  &1    & 99  &   2 & 98 &    \textit{Dis.} & \\
			\hhline{~-------}
		    \textit{} &\multicolumn{1}{|c|}{ \multirow{3}{*}{\textbf{K-means ++}}} &  \multirow{2}{*}{\textit{Single trial}}   &   95   &  5  &  97   & 3  &   \textit{Ind.} & \\
		    \cline{4-8}
   			\textit{} &	 \multicolumn{1}{|c|}{\multirow{3}{*}{\textbf{initialized centroids}}} &  & 17 &  83 & 17 &  83 & \textit{Dis.} & \\
			\hhline{|~~------}
	\textit{}  &	\multicolumn{1}{|c|}{} &  \multirow{1}{*}{\textit{Majority}}   & 98 &  2  &   100  &  0 &  \textit{Ind.}   &   \\  
			\cline{4-8}
			\begin{rotate}{90} \multirow{1}{*}{\vspace{-6pt} \textbf{(c) K-means clustering}} \end{rotate} & \multicolumn{1}{|c|}{}  & \multirow{1}{*}{\textit{voting}}   &   1 & 99    & 0     & 100  &   \textit{Dis.} & \\
			\hhline{|~-------}
		 \textit{} & \multicolumn{1}{|c}{\multirow{1}{*}{\textbf{  Hierarchical clustering}}}  &     &97  &3   & \multicolumn{1}{|c|}{95} &  5 & \textit{Ind.} & \\ \cline{4-8}
		&	 \multicolumn{1}{|c}{\textbf{initialized centroids}} &   & 16  &  84  & \multicolumn{1}{c|}{5} & 95  & \textit{Dis.}  &  \\ \hline
		\end{tabular}}
		\caption{\textbf{Confusion matrix for different clustering techniques and fixed unitary evolution [tuning stage, step (i)]}. Success percentages of the compatibility tests for all the different clustering techniques studied, i.e. \textit{bubble clustering}, \textit{Hierarchical clustering} and \textit{K-means clustering}. The latter algorithm was investigated in its standard version, and initialized by \textit{K-means++} or a preliminary run of \textit{Hierarchical clustering} \cite{NoteKMeans}. Then, majority voting was performed on the non-deterministic versions of \textit{K-means}. The reported success percentages were evaluated through numerical simulations by keeping the unitary evolution operator fixed. This choice is motivated by the need of tuning the different algorithms in order to subsequently classify new data sets.}
	\label{tab:table1}
\end{table}

(i) In the \textit{tuning stage}, we looked for the most effective clustering algorithm for our validation protocol. We are not yet applying it to validate Boson Sampling data; rather, we are optimizing the set of hyper-parameters that will define its operation (see Section \ref{section_tuning} and Supplemental Material). Indeed our protocol is based on unsupervised algorithms that, as such, do not need data with different labels to learn effective patterns.

To this aim, we applied all algorithms on numerically generated samples of output states, belonging to the collision free subspace of $n=3$ photons evolved through a fixed unitary transformation $U$ with $m=13$ modes. Hence, the dimension of the Hilbert space in this case is $\binom{13}{3}$= 286. Each algorithm was run several times, while varying the number of events within the tested samples. For each sample size, the hyper-parameters proper of each technique were optimized. 
To evaluate the success percentages for each configuration of hyper-parameters, we numerically generated $100$ distinct data sets made of three samples: two of them are drawn from the Boson Sampling distribution, while a third is drawn from the output probability distribution obtained when distinguishable particles are evolved with the same input state and unitary transformation $U$. We have performed two compatibility tests for each data set: the first between two compatible samples and the second between two incompatible ones. The results of this analysis for samples with $500$ output events are shown in Tab. \ref{tab:table1}. We observe that the best success percentage is obtained for the \textit{K-means++} method with majority voting and employing the $L_{2}$ distance. The reason for which \textit{K-means} is outperforming \textit{bubble clustering} lies in its learning capability. Indeed, due to its convergence properties through the iterations, \textit{K-means} gradually improves its insight into the internal structure that characterizes the data. This feature enables a better discrimination between compatible and incompatible samples (See Supplemental Material).

	\begin{table}[t!]
		\centering
            \footnotesize
			{\renewcommand\arraystretch{1.45}
			\begin{tabular}{|c|c|c|c||c|c|c|}
					
			\multicolumn{2}{c}{}       & \multicolumn{4}{c}{\textbf{Output Classification}}          &  \multicolumn{1}{c}{}    \\
				\hhline{~~----~}
				\multicolumn{2}{c}{}      & \multicolumn{2}{|c||}{\textit{\textbf{1-norm}}}       & \multicolumn{2}{c|}{\textit{\textbf{2-norm}}}  &   \multicolumn{1}{c}{}       \\ 
				\hhline{~-----~}
				\multicolumn{1}{c|}{}   &                \textbf{Events}      &         \multicolumn{1}{c|}{{\multirow{1}{*}{\textit{Ind.}}}}                  &  \multicolumn{1}{c||}{{\multirow{1}{*}{\textit{Dis.}}}}       &  \multicolumn{1}{c|}{{\multirow{1}{*}{\textit{Ind.}}}}     &  \multicolumn{1}{c|}{{\multirow{1}{*}{\textit{Dis.}}}}      &      \multicolumn{1}{c}{}              \\ 
				\hhline{-|-|-----|}
				\textit{}   &   \multirow{2}{*}{500}    &     95.6 $\pm$ 2.8       &   4.4 $\pm$ 2.8    &    95.7 $\pm$ 1.7                                                 &      4.3 $\pm$ 1.7     &   \textit{Ind.}        \\ 
				\hhline{|~|~|-----|}   
				\textit{}      &      \textit{}    &   69 $\pm$ 13      &       31 $\pm$ 13   &       75 $\pm$ 14    &  25 $\pm$ 14    &    \textit{Dis.} \\   
				\hhline{|~|=|=====|}    
			\begin{rotate}{90}
			\multirow{1}{*}{\vspace{-7pt}\textbf{Bubble}} \end{rotate}  &      \multirow{2}{*}{1000}   &        95.9 $\pm$ 2.0 &             4.1 $\pm$ 2.0      &       93.1 $\pm$ 2.8                                                 &      6.9 $\pm$ 2.8     & \textit{Ind.}                    \\
				\hhline{~|~|-----|}  
			\textit{} &      \textit{}         & 62 $\pm$ 30    &    38 $\pm$ 30           &   51 $\pm$ 23      &     49 $\pm$ 23   &\textit{Dis.}         \\ 
				\hhline{=======}  
			 \textit{}  &      \multirow{2}{*}{500}  &     99.1 $\pm$ 1.2      &   0.9 $\pm$ 1.2    &   99.70 $\pm$ 0.57                                             &  0.30 $\pm$ 0.57   &\textit{Ind.}        \\ 
				\hhline{|~|~|-----|}   
			\textit{}         &    \textit{}   &     45 $\pm$ 23    &   55 $\pm$ 23   &   66 $\pm$ 22   &    34 $\pm$ 22    &    \textit{Dis.}          \\   
				\hhline{|~|=|=====}   
			\multicolumn{1}{|c|}{\multirow{2}{7pt}{\textit{}}} &                                                    
			\multirow{2}{*}{1000}   &      98.7 $\pm$ 2.7  &  1.3 $\pm$ 2.7    &  96.2 $\pm$ 3.9     &  3.8 $\pm$ 3.9   & \textit{Ind.}        \\
				\hhline{|~|~|-----|}  
				\begin{rotate}{90}
				\multirow{1}{*}{\vspace{-7pt}\textbf{Kmeans++ m.v.}} \end{rotate}   &      	\textit{}     &   3.6 $\pm$ 6.4 & 96.4 $\pm$ 6.4        &      0.30 $\pm$ 0.73        &          99.70 $\pm$ 0.73   & \textit{Dis.}                    \\ 
				\hhline{|-|------|}  
			\end{tabular}}
			\caption{\textbf{Confusion matrix for \textit{bubble clustering} and \textit{K-means++} with majority voting random unitary evolution [cross-validation stage, step (ii.a)]}. Success percentages of the compatibility test for \textit{bubble clustering} and \textit{K-means} initialized with \textit{K-means++} and majority voting. These percentages were evaluated through numerical simulations, by drawing $20$ Haar-random unitary transformation, and by adopting the same hyper-parameters obtained from stage (i) corresponding to the results of Tab. \ref{tab:table1}.}
		\label{tab:table2}
	\end{table}

(ii) In the \textit{cross-validation stage}, we cross-validated the algorithm for random unitary transformations with fixed size (ii.a) and for increasing dimension of the Hilbert space (ii.b). As a first step (ii.a) we performed the test with $n=3$ photons evolving through $20$ Haar-random transformations with $m=13$ modes. For each transformation, we performed 100 tests between compatible samples and 100 between incompatible ones, by fixing the number of clusters and trials to the values determined in stage (i). In Tab. III, we report the means and standard deviations of the success percentages for a sample size of $1000$ events, and compare the obtained values  with the ones characterizing the \textit{bubble clustering} method. We observe that the chosen approach, \textit{K-means++} with majority voting and $L_{2}$ distance, indeed permits to achieve better success percentages.

\begin{table*}[ht!]
	\centering
	\includegraphics[width=\textwidth]{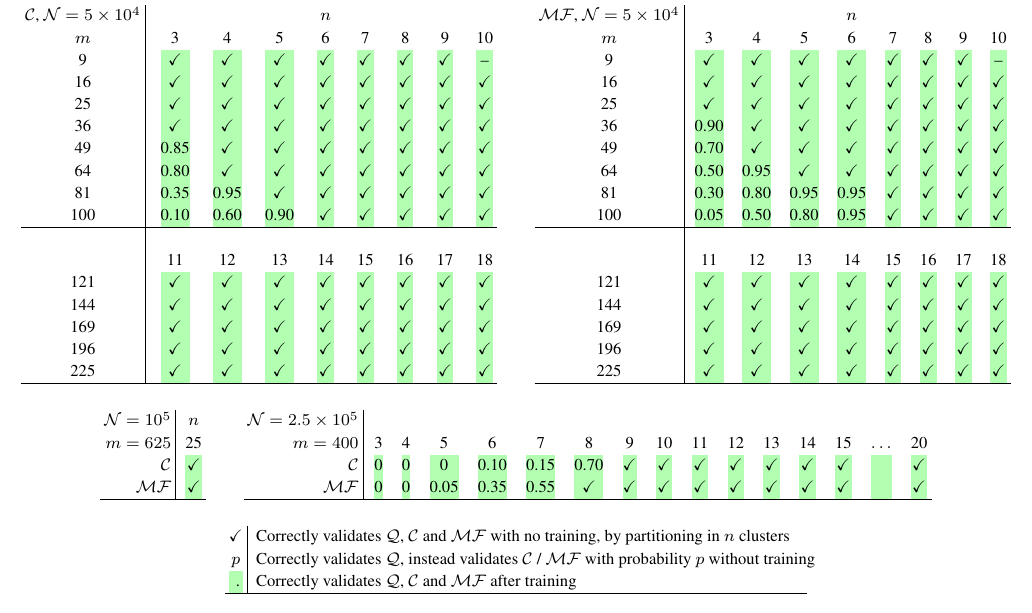}
\caption{\textbf{Efficacy of \textit{K-means++} for large-size Boson Sampling [cross-validation stage, step (ii.b)]}. The algorithm is highly effective to discern quantum ($\mathcal{Q}$) Boson Samplers from classical ($\mathcal{C}$) and Mean-Field states ($\mathcal{MF}$), even for large number of photons $n$ and modes $m$ and using very small sample sizes ($N$) as compared to the number of output combinations. For all probed ($n$, $m$), \textit{K-means} correctly identified the nature of the $Q$ sample while, when tested with adversarial samples, it still correctly identifies all their instances after proper training. \textit{K-means} is initialized by \textit{K-means++}, with optimized hyper-parameters and majority voting. Numerical samples of \textit{bona fide} Boson Samplers were generated using the algorithm by Clifford and Clifford \cite{Clifford18}. }
\label{tab:table4}
\end{table*}

To extend the cross-validation to larger-dimensional Hilbert spaces (ii.b), we exploited an efficient algorithm to sample from the distribution with distinguishable photons \cite{aaronson2013bosonsampling} and a recent algorithm by Clifford and Clifford \cite{Clifford18} to sample indistinguishable photons, with a much more efficient approach compared to the brute-force one. With this approach, we were able to test the efficacy of our protocol up to $n=25$ photons in $m=625$ modes (see Table \ref{tab:table4}). Note that, in this case, to validate a sample we require only $10^5$ events, a negligible fraction ($10^{-40}$) of the total output states. In the case $m \sim n^2$, while results are shown with fixed $N \sim 5 \times 10^{4}$ events for the sake of clarity and to avoid biases, in most instances $N \sim 10^4$ events were already sufficient. An aspect of our test that is worth noticing, as shown in Table \ref{tab:table2}, is that the probability of error is lopsided. This is a feature that can be valuable for applications where falsely concluding that trustworthy Boson Samplers are unreliable is less desirable than the converse. 
Another crucial point is that we did not need to perform a tuning of the hyper-parameters for each pair ($n$,$m$). We will discuss more in detail this aspect in Section \ref{section_tuning}.

During the cross-validation stage, we have also performed numerical simulations to verify whether the present approach is effective against other possible failure modes different from distinguishable particles, namely the Mean-Field sampler \cite{Tichy2013} and a Uniform sampler (see Section 4 of Supplemental Material). The former performs sampling from a suitable tailored single-particle distribution which reproduces same features of multiphoton interference, while the latter performs sampling from a uniform distribution. We observe that the test is capable to distinguish between a \textit{bona fide} Boson Sampler and a Uniform or Mean-Field sampler. This highlights a striking feature of this algorithm, namely the ability of our algorithm to generalize beyond the training set of distinguishable and indistinguishable samples used to learn the hyper-parameters, into situations where the data come from approximations to the Boson sampling distribution that prima facie bear no resemblance to the initial training examples.

\section{Scalable tuning of the hyper-parameter}
\label{section_tuning}

To increase the applicability of the compatibility test, we can choose suitable sets of hyper-parameters for the embedded clustering algorithms. This preliminary stage, typical in machine learning, may require proper methods such as grid search or randomized search and is in general very beneficial \cite{murphy2012}. 
In the case of the clustering algorithms described in Section \ref{numerical_results}, one common hyper-parameter is related to the minimum number of elements (sampled output events) assigned to any cluster. Specifically, \textit{Bubble clustering}, \textit{Hierarchical clustering} and \textit{K-means} require to set respectively the minimum cut-off radius, the maximum acceptable fraction of outliers (events belonging to clusters with less than $N$ elements) and the number of clusters $K$. Also, these algorithms can be applied with different notions of distance, potentially beyond the $L_{1}$ and $L_{2}$ already discussed, to reflect different knowledge on the character of a system. In this section we will clarify how to best configure the protocol to operate in instances of large dimensionality, where no algorithm for classical simulations is available to probe its functioning.

\begin{figure}[t!]
	\centering
	\includegraphics[width=\linewidth]{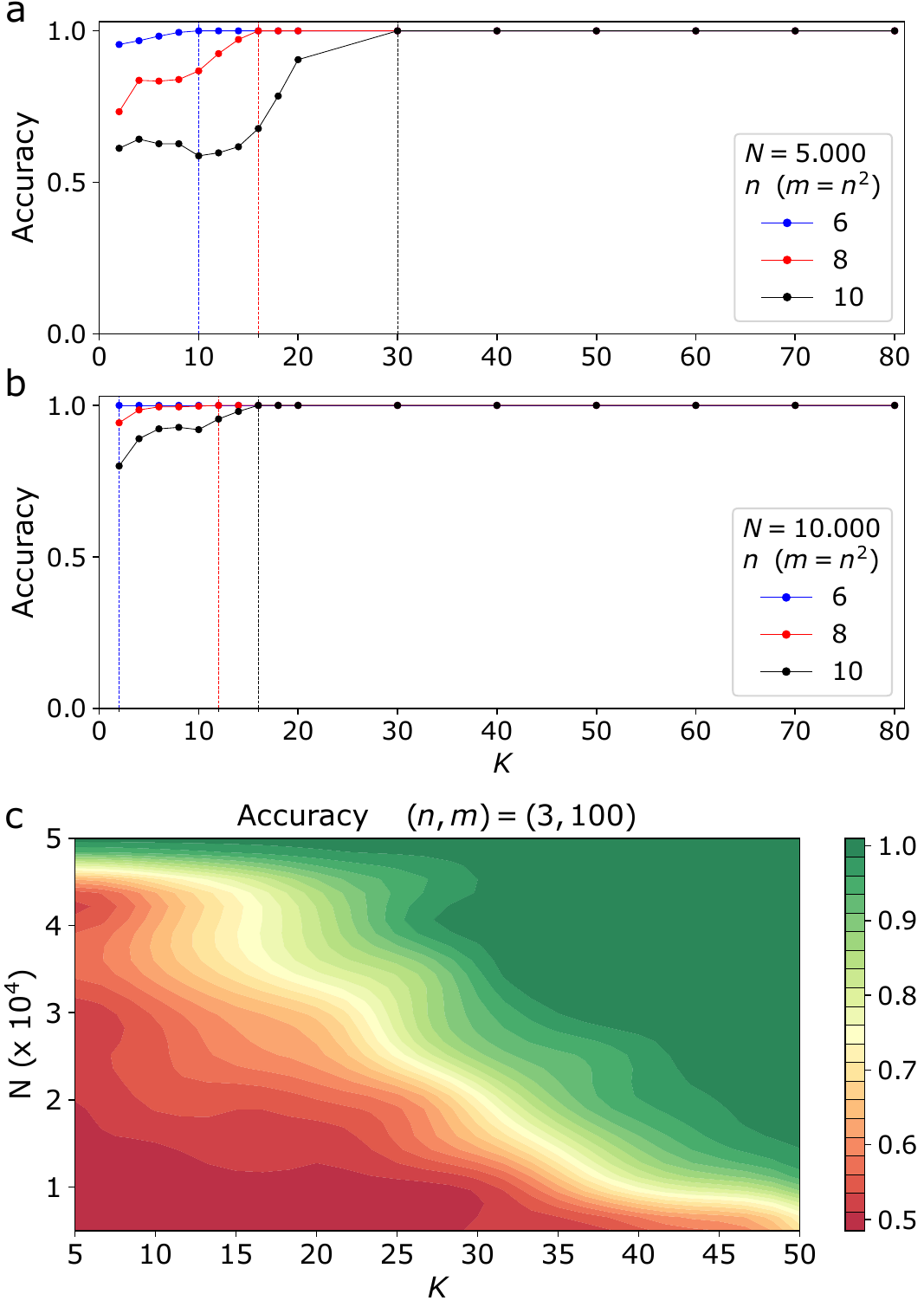}
\caption{\textbf{Test accuracy vs number of clusters and sample size}.  The compatibility test based on \textit{K-means} improves its accuracy (ratio of correct assessments) with the number of samples $N$ and clusters $K$. Hence, instead of fine-tuning $K$ for each ($n$,$m$) we can directly set the hyper-parameter to a large value, say $\frac{m}{2}\le K \le m$, with negligible computational overhead \cite{lloyd:km}. This feature has been investigated by numerically sampling $N=5 \times 10^3$ (a) or $N=10^4$ (b) photonic states with $n=6,8,10$ photons in $m=n^2$ modes. Accuracies are estimated by applying the test to numerically simulated experiments with both indistinguishable and distinguishable photons from 200 Haar-random unitary transformations. Points are connected for the sake of clarity. c) Contour plot for the accuracy in excluding classical Boson Sampling in the harder instance of $n=3$ distinguishable photons in $m=100$ modes, for different values of $K$ and $N$.  Colors describe the efficacy of the test, from red (poor) to green (perfect). All tests are performed considering a significance level of 5$\%$ for the $\chi^{2}$ test and using \textit{K-means}.}
\label{fig:scaling}
\end{figure}

In the following, let us then focus on the $L_{2}$ distance (see Section \ref{numerical_results}) and on the number of clusters $K$ for \textit{K-means}, which we have identified as the most effective technique for our purpose. We quantify the performance of the compatibility test with its accuracy, namely the success probability in ruling out samples that are not compatible with quantum Boson Sampling. In particular, we study how the choice of $K$ influences the test in two different scenarios: ($i$) when $m \sim n^2$ (Fig. \ref{fig:scaling}a), and ($ii$) for the specific instance of $(n, m)=(3,100)$, among the hardest ones presented in Table \ref{tab:table4} with $m\gg n$ (Fig. \ref{fig:scaling}b). Indeed, in the latter case the probability of bunching is practically negligible and the distributions with distinguishable and indistinguishable photons become much harder to discern for \textit{K-means}.
From this analysis, we observe that the accuracy increases with the number of samples $N$, as expected, as well as with $K$. Indeed, the observation supports the intuition that a larger $N$ provides more information to the algorithm to understand the spatial distribution in the Hilbert space, while a larger $K$ allows to probe it more finely. In particular, $K$ should be sufficiently large to appreciate the detailed spatial dishomogeneities in the Hilbert space. Moreover, smaller values of $K$ imply larger and more populated clusters that tend to average local fluctuations, so that less reliable evidence can be drawn from the compatibility test. Naturally, we also recall that $K$ cannot be chosen overly large since, in that case, each cluster would contain not enough points for the $\chi^{2}$ test to make robust predictions.
Thus, provided that the number of samples and clusters increases accordingly, the protocol is effective even in the non-favorable condition $(n, m)=(3,100)$ (see Table \ref{tab:table4} and Fig. \ref{fig:scaling}b). In the regime where $m \sim n$ or $m \sim n^2$ (Fig. \ref{fig:scaling}a) the algorithm is instead successful for almost all choices of $K$ when $N \sim 10^{4} - 10^{5}$, again in accordance with Table \ref{tab:table4}. As a general rule of thumb, which proves effective in all combinations ($n$,$m$) investigated, we set $\frac{m}{2}\le K \le m$ to satisfy the need for a bounded-above value that grows with the size of the problem. Anyway, the ultimate relevance of this specific choice can be further relaxed by collecting a larger number of samples. In this sense, the analysis reported in Fig. \ref{fig:scaling} removes the burden of a fine tuning at low dimensions, thus extending the applicability of the protocol. Moreover, the whole approach gains much also in terms of simplicity, a feature that can prove beneficial for practical applications besides the assessment of quantum supremacy.

\section{Experimental results}

\begin{figure*}[ht!]
	\centering
	\includegraphics[width=\textwidth]{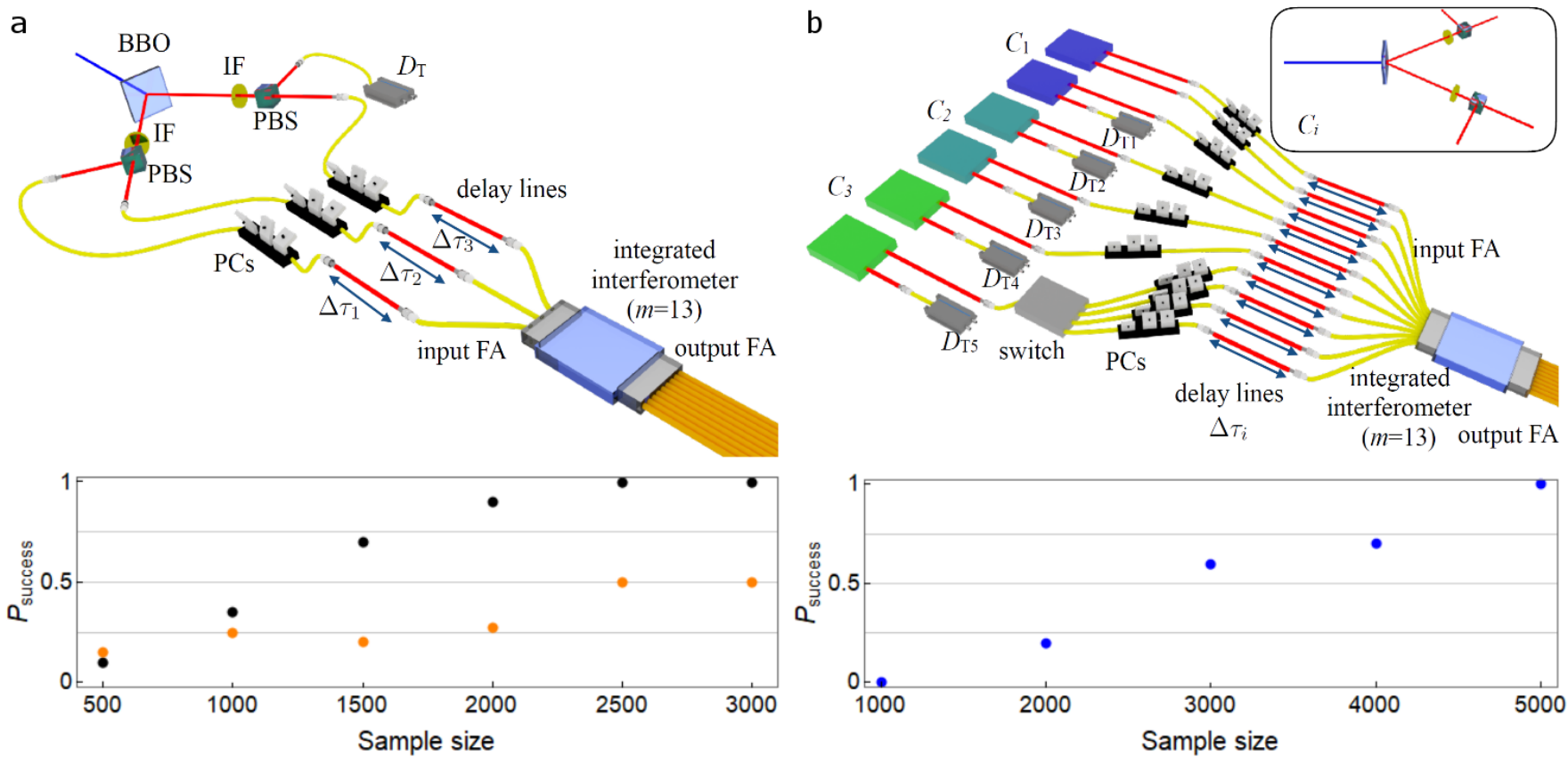}
	\caption{{\bf Experimental validation of Boson Sampling experiments}. Experimental setups for a $n=3$ standard (a) and scattershot (b) Boson Sampling experiments in an integrated $m=13$ interferometer (see Supplemental Material). Bottom insets: corresponding success probabilities of the compatibility test between inputs with indistinguishable and distinguishable photons for different sample sizes. {\bf a} Black  and orange dots refer respectively to \textit{K-means++} with majority voting and \textit{bubble clustering}. The discrepancy from the numerical results of Tab. \ref{tab:table2} is due to the non-ideal indistinguibility of the injected photons \cite{P_ind}.   {\bf b}  Input states are generated probabilistically by six independent parametric down-conversion sources (represented as boxes) in 3 different BBO crystals $C_{i}$ (see inset). Experimental data correspond to 8 different inputs. The number of events for all input states randomly varies for each sample size drawn from the complete data set. The clustering algorithm is \textit{K-means}, inizialized by \textit{K-means++}, with majority voting.  In both panels: BBO - beta barium borate crystal; IF - interferential filter; PBS - polarizing beam-splitter; PC - polarization controller; FA - fiber array. Tests are performed with a significance level of 5$\%$ and using $d=L_{2}$. }
	\label{fig:exp_BS_SCS}
\end{figure*}

Through the experimental apparatus shown in Fig. \ref{fig:exp_BS_SCS}a, we collected samples corresponding to the Boson Sampling distribution with indistinguishable and distinguishable particles. The degree of distinguishability between the input photons is adjusted by modifying their relative arrival times through delay lines (see Supplemental Material). The unitary evolution is implemented by an integrated photonic chip realized exploiting the 3D-geometry capability of femtosecond laser writing \cite{gattass2008flm} and performs the same transformation $U$ employed for the numerical results of Tab. \ref{tab:table1}. We then performed the same compatibility tests described previously on experimental data sets with different sizes, by using two methods: \textit{K-means++} with majority voting and \textit{bubble clustering}, both with 2-norm distance. The results are shown in Fig. \ref{fig:exp_BS_SCS}a, for the case of incompatible samples. This implies that the reported percentages represent the capability of the test to recognize Boson Sampler fed with distinguishable photon inputs.
Reshuffling of the experimental data was used to have a sufficient number of samples to evaluate the success percentages (see Supplemental Material). Hence, the tests were performed on samples drawn randomly from the experimental data.

\section{Generalization\\ for scattershot Boson Sampling} 
The scattershot version of Boson Sampling \cite{lund2014, scattershot2015} is implemented through the setup of Fig. \ref{fig:exp_BS_SCS}b. Six independent parametric down-conversion photon pair sources are connected to different input modes of the $13$-mode integrated interferometer. In this case, two input modes (6,8) are always fed with a single photon. The third photon is injected probabilistically into a variable mode, and the input is identified by the detection of the twin photon at trigger detector T$_{i}$. We considered a generalization of the proposed algorithm to be applied for scattershot Boson Sampling. In this variable-input scenario a Boson Sampler to be validated provides $N$ samples that correspond to $N$ different inputs of the unitary transformation, that is, $N$ Fock states $\Phi_{i}$ with $i \in \{1, n\}$. Hence, our validation algorithm in its standard version needs to perform $N$ separate compatibility tests. Indeed, it would bring $N$ distinct chi-square variables $\chi^{2}_{i}$, where the $i$-th variable would quantify the agreement between the distribution of the data belonging to the input $\Phi_{i}$ and the distribution of a sample drawn by a trusted Boson Sampler with the same input state. Hence, each input state would be tested separately.

In order to extract only one quantity to tell whether the full data set is validated or not, for all inputs, a new variable can be defined as $\tilde{\chi}^2=\sum_{i=1}^{N}\chi^{2}_{i}$. This variable is a chi-square one with $\nu= \sum_{i=1}^{N} \nu_{i}$ degrees of freedom, provided that the $\chi^{2}_{i}$ are independent. We have performed this generalized test on the experimental data by adopting the same clustering technique previously discussed in the single input case. 

\section{Experimental results\\ for scattershot Boson Sampling} 
We have collected output samples given by 8 different inputs both with indistinguishable photons and with distinguishable ones. Through the evaluation of the new variable $\tilde{\chi}^2$, the algorithm was able to distinguish between a trustworthy scattershot Boson Sampler and a fake one at the significance level of 5$\%$, using a total number of observed events up to $5000$ events (over all inputs), as shown in Fig.\ref{fig:exp_BS_SCS}b. The standard version of the test, validating each input separately, would require samples of 2000 events per input to reach a success percentage $\geq 80\%$, that is, an overall amount of $16000$ events. Hence, the generalized version of the test permits to significantly reduce the amount of necessary resource to validate scattershot Boson Sampling experiments.

\section{Structure of the probability distributions} 
Our previous discussion has conclusively shown that, at least for the values of $(n,m)$ considered, $K$--means clustering algorithms are highly effective at discriminating boson samplers that use distinguishable photons versus those with indistinguishable ones.  Here we provide further analysis that shows why our approach is effective at this task and sheds light on how future tests could be devised to characterize faulty boson samplers.
We address this aspect by providing numerical evidence to explain the physical mechanism behind the correct functioning of our validation test. 

The clustering techniques that form the basis of our pattern recognition methodology rely on aggregating the experimental data according to the distance between the output states. The key observation is that the number of events necessary to effectively discriminate the samples is dramatically lower than the number of available output combinations. In particular, such fraction drops fast to smaller values by increasing the system size $(n,m)$. For instance, $10^4$ events would correspond to $0.08$ of the Hilbert space dimension for $(4,40)$, to $10^{-10}$ for $(10,100)$ and to only $10^{-30}$ for $(20,400)$. Accordingly the output sample from the device will mostly consist of output states occurring with no repetition. Hence, only the configurations presenting higher probability will effectively contribute to the validation test. 

For the sake of clarity let us focus on the discrimination between indistinguishable and distinguishable particles.  We leave for subsequent work the task of explaining why other alternative models, such as Mean-Field states, are also noticed by our approach.  More specifically, we analyze the structure of the outcome distributions for the two cases. Since data clustering is performed according to the distance between states, the method can be effective if (i) the distributions of the output states exhibit an internal structure and (ii) correlations between distributions with different particle types are low.

\begin{figure*}[ht!]
\centering
\includegraphics[width=0.99\textwidth]{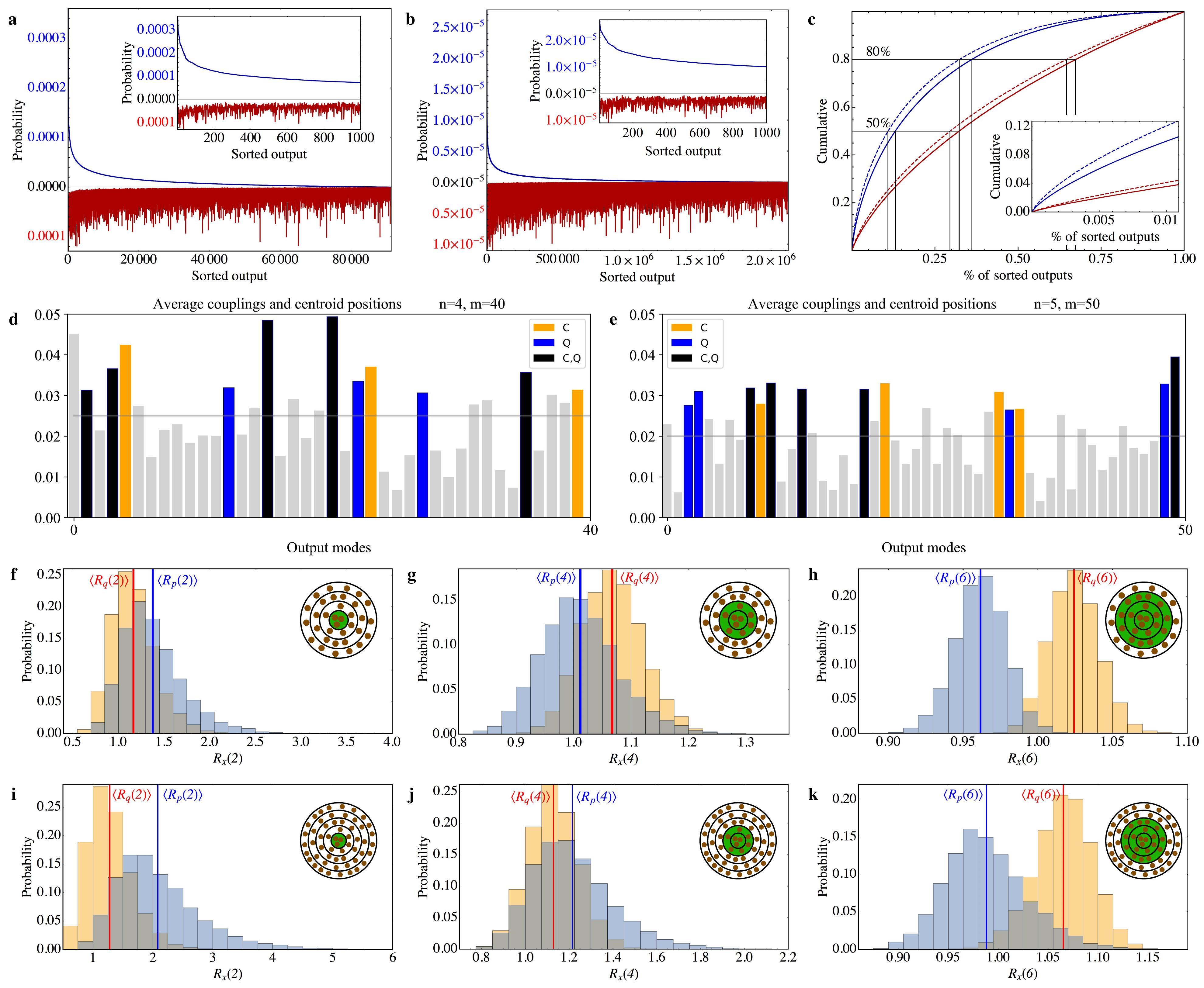}
\caption{{\bf Analysis of the structure of the distributions}. {\bf a,b}, Probability distributions for a fixed unitary $U$ in the case of indistinguishable (blue) and distinguishable (red) photons. {\bf a}, $n=4$, $m=40$, and {\bf b}, $n=5$, $m=50$. The distributions are sorted by following the same ordering, so as to have decreasing probability values for the indistinguishable distribution $P_{i}$. Inset: zoom corresponding to the 1000 most probable output states. {\bf c}, Cumulative distributions for indistinguishable (blue) and distinguishable (red) photons, by following the same ordering of panels {\bf a,b}. Solid lines: $n=4$, $m=40$. Dashed lines: $n=5$, $m=50$. Black lines highlight the levels corresponding to $50\%$ and $80\%$ of the overall probability, which require approximately twice the number of output states in the distinguishable case. Inset: zoom corresponding to $0.01\%$ most probable output states. {\bf d-e}, In the $m$-dimensional vector space where \textit{K-means} is performed on quantum (Q) and classical (C) samples, centroids tend to be located where the output modes have high probability averaged over the input modes. This property is signalled by a vector with one coordinate, out of $m$ for a Fock state, significantly higher than the others. Zeroing the small ones out yields a plot analogous to the one shown here for 10 clusters with ({\bf d}) $n=4$, $m=40$ and ({\bf e}) $n=5$, $m=50$. {\bf f-k}, Histograms of the ratios $R_{p}(k)$ (cyan) and $R_{q}(k)$ (orange) between the overall probability included within a sphere of $L_{1}$-norm $\leq k$. {\bf f,h}, $n=4$, $m=40$. {\bf i-k}, $n=5$, $m=50$. Vertical lines correspond to the averages $\langle R_{x}(k) \rangle$, with $x=p$ (blue) and $x=q$ (red). {\bf f,i}, $k=2$, {\bf g,j}, $k=4$ and {\bf h,k}, $k=6$. Insets: schematic view of the spheres at distance $\leq k$, represented by concentric circles, where states are represented by brown points.}
\label{sorting}
\end{figure*}

As first step towards this goal, we computed the probability distributions with $n=4$ indistinguishable photons $(P_{j})$ and distinguishable particles $(Q_{j})$, for a fixed unitary transformation $U$ with $m=40$ modes. Fig. \ref{sorting}a reports the two distributions sorted according to the following procedure, in order to highlight their different internal structure. The distribution with indistinguishable photons is sorted in decreasing order starting from the highest probability, while the distribution with distinguishable particles is sorted by following the same order adopted for the indistinguishable case. More specifically, the first element is the value of $Q_{j}$ for the output state corresponding to the highest value of $P_{j}$, the second element corresponds to the state with the second highest value of $P_{j}$, and analogously for all other terms. We observe a small correlation between the $P_j$ and $Q_j$ distributions. To quantify this feature, we computed two different statistical coefficients (the Pearson $r$ and the Spearman's rank $\rho$ ones), that are employed to evaluate the presence of linear or generic correlations between two random variables. In particular we find that the Pearson correlation coefficient is $r \sim 0.56$, while the Spearman's rank coefficient is $\rho \sim 0.55$, which suggest that the two distributions have different supports over the outcome distributions. The same analysis have been performed for $n=5$ and $m=50$, showing that a similar behavior is obtained for increasing size (see Fig. \ref{sorting}b). By averaging over $M^{\prime}=100$ different unitaries, the correlation coefficients are $r \sim 0.62 \pm 0.03$ and $\rho \sim 0.64 \pm 0.04$ (1 standard deviation) for $n=4$ and $m=40$, and $r \sim 0.57 \pm 0.03$ and $\rho \sim 0.62 \pm 0.04$ (1 standard deviation) for $n=5$ and $m=50$. These results show that the low values of the correlations between $P_{j}$ and $Q_{j}$ do not depend on the specific transformation $U$, and that this behavior is maintained for larger size systems. Similar conclusions are observed in the cumulative distributions (see Fig.~\ref{sorting}c), where the distinguishable case is sorted by following the same order of the indistinguishable one. We observe that, for the cumulative probability for distinguishable bosons to reach the same value attained for indistinguishable bosons, a significantly larger portion of the Hilbert space has to be included. For instance, when $n=4$ and $m=40$, $50\%$ of the overall probability is achieved by using $\sim 13\%$ of the overall number of outputs for indistinguishable photons, while $\sim 32\%$ are necessary for the distinguishable case (by following the above mentioned ordering procedure). Similar numbers are obtained for larger dimensionalities ($\sim 11 \%$ and $\sim 30 \%$ respectively when $n=5$ and $m=50$).

The second crucial aspect of our method is related to the localization of outcomes with the highest probabilities. More specifically, this approach can be effective in constructing useful cluster structures if the most probable states are surrounded by other states with high probability. In this way, when a number of events much lower that the number of combinations is collected, the outcomes actually occurring in the data sample will present lower distance values thus justifying the application of a clustering procedure. An intuition for this feature is provided in Fig. \ref{sorting}d,e, which shows that centroids tend to locate in the positions of the $m$-dimensional vector space corresponding to the output modes with the highest probability, averaged over the input modes.

We further probe how these correlations become visible through a clustering method by we performing numerical simulations that randomly vary the unitary transformation $U$ for ($n=4$, $m=40$) and ($n=5$, $m=50$). For each sampled transformation $U$, we calculated the probabilities $P_{j}$ and $Q_{j}$ for both cases (indistinguishable and distinguishable photons) and then sorted the distribution $P_{j}$ in decreasing order. Let us call $J$ the outcome with the highest $P_{j}$ value which is to say $J = {\rm argmax}(P_j)$. Let us for simplicity fix the distance to be the $L_{1}$-norm (analogous results are obtained for the $L_{2}$-norm). Note that the $L_{1}$-norm defined in the main text has only $N$ possible non-trivial values $k=2 s$, with $s=1,\ldots n$. We then estimated the overall probability $P(k) = \sum_{j:\|j-J\|_1\leq k} P_{j}$, where $P(k)$ is the probability included in a sphere with distance $\leq k$ computed using the $L_1$ norm. The same calculation is performed for the distinguishable particle case $Q(k) = \sum_{j:\|j-J\|_1\leq k} Q_{j}$, by using the same outcome value $J$ as a reference. 

We study the ratio $R_{p}(k)=P(k)/Q(k)$ between the two probabilities, that can be thought of as a likelihood ratio test wherein $R_p(k)>1$ implies that the evidence is in favor of indistinguishable particles and conversely $R_p(k)<1$ suggests that the particles are distinguishable.  Such comparison is then performed for $M^{\prime \prime}=100$ different unitary matrices $U$ and by using as reference outcome $J$ the $M_{\mathrm{max}}=100$ highest probability outcomes for each $U$.  The results are reported in Fig. \ref{sorting}d-f for $n=4$, $m=40$ with $k=2,4,6$ (being $k=8$ a trivial one, which includes all output states given $4$-photon input states). The analysis is also repeated in the opposite case, where the data are sorted according to the distinguishable particle distribution $Q_{j}$ and $R_q = Q(k)/P(k)$. We observe that $R_{p}(2)$ has an average of $\langle R_{p}(2) \rangle \sim 1.4$, and that $P(R_{p}(2)>1) \sim 0.904$. For increasing values of $k$, $\langle R_{p}(k) \rangle$ converges to unity since a progressively larger portion of the Hilbert space is included thus converging to $R_{p}(k=8)=1$ (respectively, $\sim 0.16 \%$ for $k=2$, $\sim 4.3 \%$ for $k=4$ and $\sim 35.5\%$ for $k=6$). Similar results have been obtained also for $n=5$ and $m=50$ (see Fig. \ref{sorting}g-j), where $\langle R_{p}(2) \rangle \sim 2.08$, and that $P(R_{p}(2)>1) \sim 0.986$. This behavior for $R_{p}(k)$ and $R_{q}(k)$ highlights a hidden correlation within the output states distributions, where outcomes with higher probabilities tend to be more strongly localized at low $L_1$ distance from the reference outcome for indistinguishable bosons than those drawn from a distribution over distinguishable particles. This is why basing an algorithm on this feature is effective for diagnosing a faulty boson sampler that uses distinguishable photons.

The same considerations can be obtained also from a different perspective. Indeed, it has been recently shown \cite{Walschaers2016} that information on particle statistics from a multiparticle experiment can be retrieved by low-order correlation measurements of $C_{ij} = \langle n_{i} n_{j} \rangle - \langle n_{i} \rangle \langle n_{j} \rangle$, where $n_{i}$ is the number operator. Correlations between the states of the output distribution, originating from the submatrix of $U$ that determines all output probabilities, will correspond to correlations between the output modes. Such correlations are different depending on the particle statistics (indistinguishable or distinguishable particles) due to interference effects, and can thus be exploited to identify the particle type given an output data sample. More specifically, a difference between particle types is observed in the moments of the $C_{ij}$ set, thus highlighting different structures in the output distributions. As previously discussed, such different structures can be detected by clustering approaches.

To summarize, all these analyses show that Boson Sampling distributions with indistinguishable and distinguishable particles present an internal structure that can be catched by the clustering procedure at the basis of our validation method, thus rendering our method effective to discriminate between the two hypotheses.

\section{Discussion} 
In this article we have shown that pattern recognition techniques can be exploited to identify pathologies in Boson Sampling experiments.  The main feature of the devised approach relies in the absence of any permanent evaluation, thus not requiring the calculation of hard-to-compute quantities during the process. The efficacy of these pattern recognition techniques relies on the presence of marked correlations in the output distributions, that are related to the localization of the outcomes with the highest probabilities and that depend on the particle type. Additionally, the absence of any assumptions on the system under study allows to apply the compatibility test to a much broader class of multi-particle states: one would just need a trusted sample from an arbitrary class to check whether or not a different sample is compatible with it.

This approach can be also adopted in larger Hilbert spaces with arguably no need for a fine tuning of clustering hyper-parameters, which makes it a promising approach for identifying flaws in the next generation of Boson Samplers. We can further adopt the test as part of the validation toolbox at the boundaries of quantum supremacy, where classical and quantum sampling would take approximately the same time and it is still possible to numerically generate the trusted samples.

We also envisage that other protocols, based on more sophisticated machine learning methods, might in future provide even more effective solutions to this aim. Moreover, our experimental demonstration shows that it is possible to successfully test Boson Sampling even in lossy scenarios, which have already been shown to maintain the same computational hardness of the original problem \cite{Aaronson2016}.
Looking forward, it is our hope that when building data-driven  (rather than first principles) models for error, cross-validation will be used to report the performance of such algorithms.  For example, our method had $100\%$ classification accuracy for the training data but had roughly $95\%$ accuracy in the test data.  Had we only reported the performance of the algorithm on the training data it would have provided a misleading picture of the method's performance for larger Boson Sampling experiments.  For this reason it is important that, if we are to use the tools of machine learning to help validate quantum devices, then we should also follow the lessons of machine learning when reporting our results.

While our work shows that machine learning can be used to provide evidence that a boson sampler is faulty, it does not provide a definitive test.  Furthermore, even if a boson sampler passes such tests it need not also be a valid boson sampler.  This means that while machine learning is a valuable tool to help build confidence in boson samplers, it does not solve the validation problem in and of itself.  Finding ways to clearly state the assumptions under which such machine learning approaches validate a boson sampler, and the a posteriori probability with which it is found to be valid, remains an open problem.

Finally, although our work is focused on validation of Boson samplers, it is important to note that the lessons learned from this task are more generally applicable.  Unsupervised methods, such as clustering, can be used to find patterns in high-dimensional data that allow simple algorithms to learn facts about complex quantum systems that humans can easily miss. As a simple example, we showed that the centroids' positions are correlated to the modes where the single-particle probability is higher on average. By continuing to incorporate ideas from computer vision into our verification and validation toolbox we may not only develop the toolbox necessary to provide a convincing counterexample to the extended Church-Turing thesis, but also provide the means to debug the first generation of fault tolerant quantum computers.

\section*{Acknowledgments}
This work was supported by the ERC-Starting Grant 3D- QUEST (3D-Quantum Integrated Optical Simulation; Grant Agreement No. 307783), by the H2020-FETPROACT-2014 Grant QUCHIP (Quantum Simulation on a Photonic Chip; Grant Agreement No. 641039), and by the European Research Council (ERC) Advanced Grant CAPABLE (Composite integrated photonic platform by femtosecond laser micromachining, Grant Agreement No. 742745).


\begin{thebibliography}{10}
\bibitem{NoteKMeans} Though $K$-means is not guaranteed to converge for metrics other than $L_2$, $L_1$ proved equally effective in our protocol. In this case it is also possible to use K-medians, since the median is the best $L_1$ estimator just like the mean for $L_2$.
\end{thebibliography}
\end{document}


\title{Supplementary Information\\ 
Pattern recognition techiques for Boson Sampling validation}

\author{Iris Agresti}
\affiliation{Dipartimento di Fisica, Sapienza Universit\`{a} di Roma,
Piazzale Aldo Moro 5, I-00185 Roma, Italy}

\author{Niko Viggianiello}
\affiliation{Dipartimento di Fisica, Sapienza Universit\`{a} di Roma,
Piazzale Aldo Moro 5, I-00185 Roma, Italy}

\author{Fulvio Flamini}
\affiliation{Dipartimento di Fisica, Sapienza Universit\`{a} di Roma,
Piazzale Aldo Moro 5, I-00185 Roma, Italy}

\author{Nicol\`o Spagnolo}
\affiliation{Dipartimento di Fisica, Sapienza Universit\`{a} di Roma,
Piazzale Aldo Moro 5, I-00185 Roma, Italy}

\author{Andrea Crespi}
\affiliation{Istituto di Fotonica e Nanotecnologie, Consiglio
Nazionale delle Ricerche (IFN-CNR), Piazza Leonardo da Vinci, 32,
I-20133 Milano, Italy}
\affiliation{Dipartimento di Fisica, Politecnico di Milano, Piazza
Leonardo da Vinci, 32, I-20133 Milano, Italy}

\author{Roberto Osellame}
\affiliation{Istituto di Fotonica e Nanotecnologie, Consiglio
Nazionale delle Ricerche (IFN-CNR), Piazza Leonardo da Vinci, 32,
I-20133 Milano, Italy}
\affiliation{Dipartimento di Fisica, Politecnico di Milano, Piazza
Leonardo da Vinci, 32, I-20133 Milano, Italy}

\author{Nathan Wiebe}
\affiliation{Station Q Quantum Architectures and Computation Group, Microsoft Research, Redmond, WA, United States}

\author{Fabio Sciarrino}
\affiliation{Dipartimento di Fisica, Sapienza Universit\`{a} di Roma,
Piazzale Aldo Moro 5, I-00185 Roma, Italy}

\maketitle

\section*{SUPPLEMENTARY NOTE 1: EFFICIENCY OF THE DIFFERENT PATTERN RECOGNITION TECHINQUES}
Let us analyze in more detail the characteristics of two of the different pattern recognition techniques that we applied to Boson Sampling validation, i.e. \textit{K-means} and \textit{bubble clustering}. Specifically, the main strength of \textit{K-means clustering} [S1] lies in the ability of the algorithm to learn, through a certain number of iterations, a locally optimal cluster structure for the experimental data to be analyzed.
This feature is clearly highlighted when we compare the performances of the compatibility test, that is the resulting p-values, when we adopt \textit{K-means clustering} and \textit{bubble clustering}. In Supplementary Fig. \ref{fig2} a we compare two incompatible samples, i.e. one drawn from a trustworthy Boson Sampler and one from a Boson Sampler fed with a distinguishable three photon input.
In Supplementary Fig. \ref{fig2} b we instead compare two compatible samples, both drawn from a trustworthy Boson Sampler.
\textit{Bubble clustering} is not iterative and the resulting cluster structure is deterministic for a given data set. Indeed, the algorithm chooses the cluster centers accordingly to the observation frequency of the output states and assigns the elements only once. On the other hand, \textit{K-means} convergence is not univocal, since the final cluster structure depends on the centroids inizialization, and, through the selection of the centroids and the corresponding assigned elements, it is guaranteed to locally optimize its objective function, $\frac{1}{N} \sum_{j=1}^{k}\sum_{i=1}^{n_{k}} d(e_{ij},c_{j})$ [S2]. Having considered this feature of \textit{K-means}, we performed 200 tests both on the two compatible samples and on the incompatible ones, using 100 different uniformly drawn random inizializations, and 100 \textit{K-means++} inizializations.
The pattern recognition techniques based on \textit{K-means}, given its objective function convergence to local minima property, improve the test's result after each iteration, being able to recognize whether the samples are compatible or not with better results than \textit{bubble clustering}, that is, leading to a smaller p-value, with incompatible samples and a greater p-value with compatible samples. The p-value quantifies the significance of the $\chi^{2}$ test result, since it gives the probability of obtaining a greater value of the $\chi^{2}$ variable, if the null hypothesis of compatible samples is true.

\begin{suppFig}[h!]
	\centering
	\includegraphics[width=\textwidth]{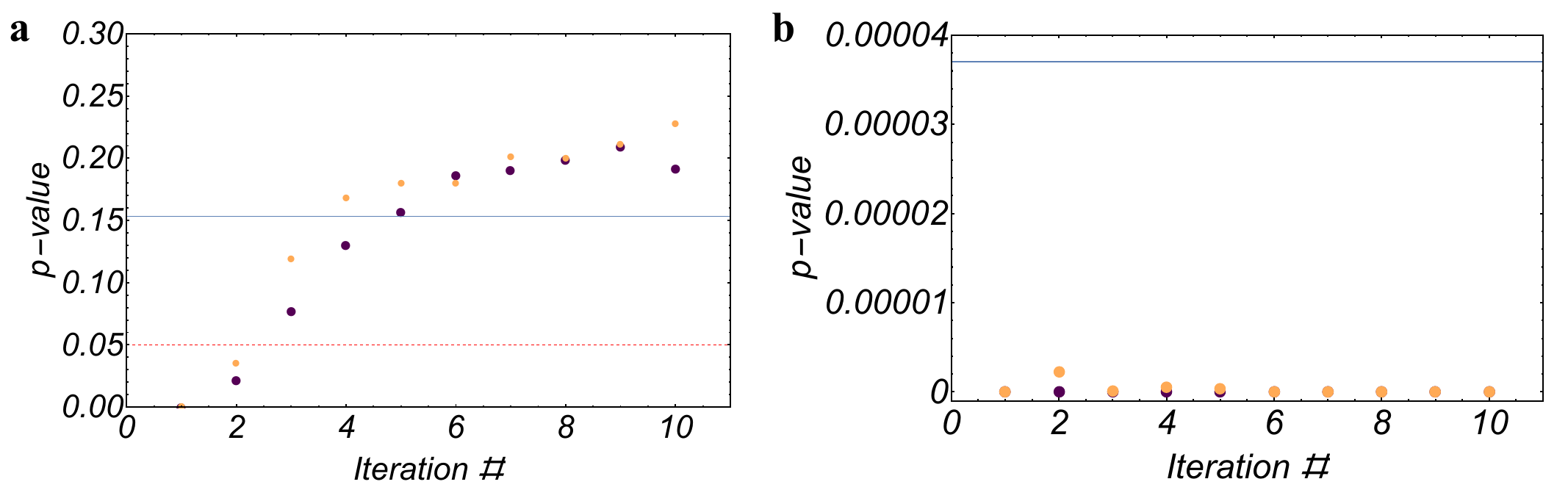}
	\caption{{\bf \textit{K-means clustering} vs \textit{bubble clustering}}. {\bf a)}, P-values obtained by the application of our compatibility test on two compatible samples of 2000 events, both  drawn from a trustworthy Boson sampler, with $N$=3 and $m$=13. The blue horizontal line indicates the result obtained by using the cluster structure given by \textit{bubble clustering}. The darker and lighter dots represent the p-values obtained respectively from the application of the test on the cluster structures obtained by \textit{K-means}, initialized by uniformly drawn random centroids and by \textit{K-means++}, at each iteration. Since the obtained cluster structure is not deterministic, we performed the mean on the p-values corresponding to 100 different inizializations. The red dashed line indicates the p-value above which the two samples are recognized as compatible.
		{\bf b)}, P-values obtained by the application of our compatibility test on two incompatible samples of 2000 events, one drawn from a trustworthy Boson sampler and the other from a Boson Sampler fed with a distinguishable photon input. The blue horizontal line indicates the result obtained via \textit{bubble clustering}. The darker and lighter dots represent, respectively, the p-values obtained by \textit{K-means}, initialized by uniformly drawn random centroids and via \textit{K-means++} inizialization. Since the cluster structure  obtained by \textit{K-means} is not deterministic, we performed the mean on the p-values corresponding to 100 different inizializations.}
	\label{fig2}
	
\end{suppFig}

\section*{SUPPLEMENTARY NOTE 2: HALTING CONDITION FOR HIERARCHICAL CLUSTERING}
The choice of the halting condition for \textit{Hierarchical clustering} attempts to balance two conflicting requirements: the number of clusters and their minimum size. Firstly, there is the need to have clusters with at least 5 elements, to make the $\chi^{2}$ test meaningful. However, the minimum size is not the optimal choice as halting condition, since there is no control over the number of clusters and there is the risk that the algorithm stops with too few clusters remaining. An excessively low number of clusters, less than 3, can compromise the correct operation of the test. We therefore chose to remove outlier elements in the data, neglecting them in the final cluster structure. This is done by requiring that the algorithm stops when a predetermined fraction of the elements belongs to clusters with less than 5 elements. The fraction was then tuned to maximize the efficiency of the validation test, while monitoring the amount of data taken away.

 	\section*{SUPPLEMENTARY NOTE 3: RESHUFFLING OF THE EXPERIMENTAL DATA}
 	Our evaluation of the experimental success percentages of the validation test uses 100 sets of three data samples, two compatible and one incompatible. 
 Having such a number of data allows to perform 100 compatibility tests between compatible samples and 100 between incompatible samples. The required sample size to ensure more than 0.8 probability to recognize two incompatible samples is 500 events. 
 	This sample size implies that, for each input mode occupation list, we would need at least 5x$10^4$ events drawn from the distribution with distinguishable photon input and 10x$10^5$ events drawn from the corresponding Boson sampling distribution.
 	Since the amount of experimental data available was not sufficient, we adopted a reshuffling of the experimental data. To this purpose, starting from a set of $N_{exp}$ output events sampled experimentally, we picked randomly the number of events corresponding to the required sample size $N_{e}$ among the available data, as long as $N_{e}$ < $N_{exp}$.
 	This approach may bring a small bias in the results. To quantify this bias, we performed a numerical simulation, comparing the trends of the success percentage as a function of the sample size, with and without reshuffling. The trend is the same, but the growth is slower. Specifically, as reported in Supplementary Figure \ref{fig3}, numerical simulations show that, for the investigated sample sizes, reshuffling of data does not affect the result of the validation algorithm adopting \textit{K-means clustering}, initialized with \textit{K-means++}. This analysis implies that the success percentages obtained experimentally and showed in Fig. 3 b and Fig. 4 of the main text are reliable.

 	 \begin{suppFig}[h!]
 		\centering
 	\includegraphics[width=\textwidth]{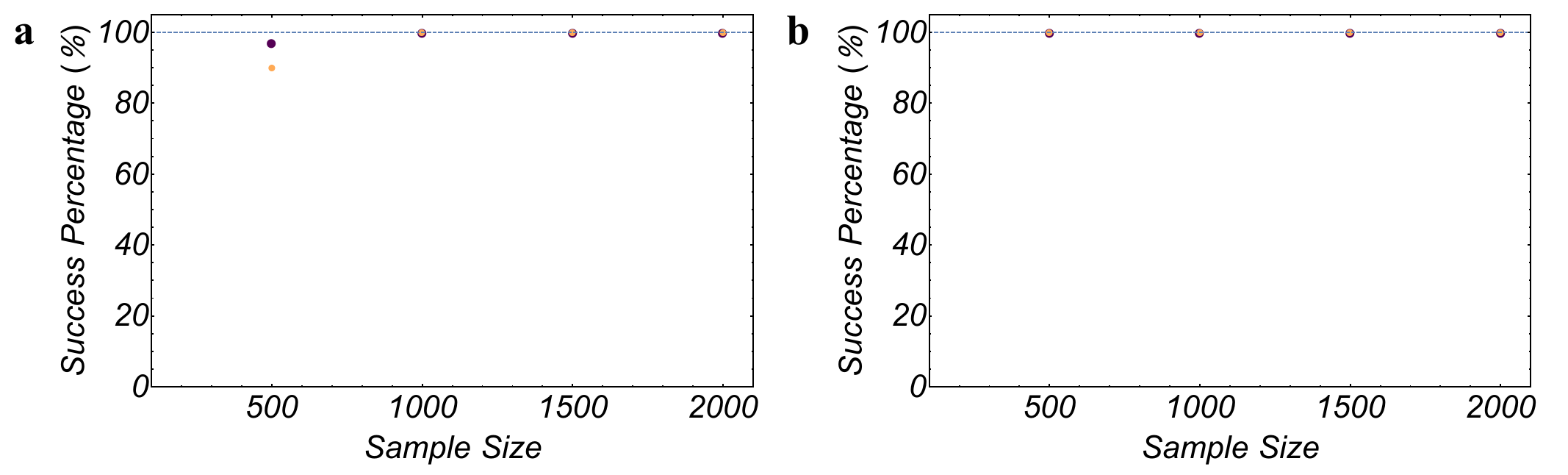}
 		\caption{{\bf Reshuffling of the data}. \textbf{a} Success Percentages of the compatibility test performed on incompatible samples, with $N$=3 and $m$=13, with \textit{K-means clustering} initialized with \textit{K-means++}. The lighter and darker dots represent the percentages corresponding respectively to reshuffled and to not reshuffled data. 
 		\textbf{b} Success Percentages of the compatibility test on incompatible samples, with $N$=3 and $m$=13, adopting \textit{K-means clustering} initialized with \textit{K-means++}, with majority voting. The lighter and darker dots represent the percentages corresponding respectively to reshuffled and to not reshuffled data. The percentages here indicated are evaluated considering 5$\%$ as significance level for the $\chi^{2}$ test.}
 		\label{fig3}
 		
 	\end{suppFig}

\section*{SUPPLEMENTARY NOTE 4: BOSON SAMPLING VS UNIFORM AND MEAN-FIELD SAMPLING WITH K-MEANS++}
   We analyzed the efficiency of our pattern recognition validation protocol also to discriminate \textit{bona fide} Boson Samplers from uniform samplers and Mean-Field samplers.
 Mean-Field sampler ($Mf$), firstly described in [S3], is a physically plausible and efficiently simulable system that reproduces many of the interference features characteristic of a genuine Boson Sampler. In this scenario, with \textit{N} bosons and \textit{m} modes in  the probability distribution is given by the  following equation:
 
 \begin{equation}
 P_{Mf}(T_{Mf}, S_{Mf}, U)= \frac{N!}{N^{N}\sum_{l=1}^{N}}\sum_{q=1}^{N}\prod_{k=1}^{N}|U_{k_{q}j_{k}}|^2
 \end{equation}
where $T_{Mf}$=($t_{1}$, ..., $t_{m}$) is the output configuration and $S_{Mf}$=($s_{1}$, ..., $s_{m}$) the input one, with respectively ($k_{1}$,..., $k_{N}$) and ($j_{1}$, ..., $j_{N}$) as mode arrangements.
   So, for each of Hilbert space dimensions shown in Tab. III of the main text, we drew 10 samples of 6000 events from the three mentioned distributions (Boson Sampling, Mean-Field and uniform distribution), for a fixed unitary transformation $U$. 
   We then compared samples drawn from the Boson Sampling distribution to those drawn from the alternative samplers, adopting the same parameters used in Tab. III of the main text (i.e. sample size of 6000 events and 25 clusters).
   As shown in Supplementary Tab. I, the test was able to recognize the three kinds of samples as incompatible, with a considered significance level of 1$\%$. Indeed, given the high performance of our algorithm in these tests it should come as no surprise that including  majority voting on top of the previous results did not yield any further advantage. Note that the test is successfully applied to distinguish between different failure modes that were not considered in the training stage.  The fact that our algorithm has never seen such proxies for the Boson sampling distribution and yet performs exceedingly well in classifying samples drawn underscores the fact that our approach yields much broader tools for validating Boson samplers than the limited scope of the training data used may suggest.

   \begin{table}[ht!]
  \begin{tabular}{c|c|c|c}
	\hhline{~~~~}
	\multicolumn{4}{c}{\multirow{2}{*}{\textbf{Output Classification}} }\\
	\hhline{~~~~}
	\multicolumn{4}{c}{} \\
	\hhline{~--~}
	\multicolumn{1}{c}{\textbf{(N, m)}} & \multicolumn{1}{|c|}{\textit{Ind.}}	& \multicolumn{1}{|c|}{\textit{M. F.}}	& \multicolumn{1}{c}{}\\
	\hhline{-==-}
	\multicolumn{1}{|c}{\multirow{2}{*}{\textbf{(3,13)}}} & \multicolumn{1}{|c|}{\textit{10}}	& \multicolumn{1}{|c|}{\textit{0}}	& \multicolumn{1}{c|}{\textit{Ind.}}\\
	\hhline{~---}
	\multicolumn{1}{|c}{\multirow{2}{*}{}} & \multicolumn{1}{|c|}{\textit{0}}	& \multicolumn{1}{|c|}{\textit{10}}	& \multicolumn{1}{c|}{\textit{M. F.}}\\
	\hhline{====}
	\multicolumn{1}{|c}{\multirow{2}{*}{\textbf{(5,50)}}} & \multicolumn{1}{|c|}{\textit{10}}	& \multicolumn{1}{|c|}{\textit{0}}	& \multicolumn{1}{c|}{\textit{Ind.}}\\
	\hhline{~---}
	\multicolumn{1}{|c}{\multirow{2}{*}{}} & \multicolumn{1}{|c|}{\textit{0}}	& \multicolumn{1}{|c|}{\textit{10}}& \multicolumn{1}{c|}{\textit{M. F.}}\\
	\hhline{====}
	\multicolumn{1}{|c}{\multirow{2}{*}{\textbf{(6,50)}}} & \multicolumn{1}{|c|}{\textit{9}}	& \multicolumn{1}{|c|}{\textit{1}}		& \multicolumn{1}{c|}{\textit{Ind.}}\\
	\hhline{~---}
	\multicolumn{1}{|c}{\multirow{2}{*}{}} & \multicolumn{1}{|c|}{\textit{0}}	& \multicolumn{1}{|c|}{\textit{10}}	& \multicolumn{1}{c|}{\textit{M. F.}}\\
	\hhline{----}
	\end{tabular}  
    \quad
   \begin{tabular}{c|c|c|c}
		\hhline{~~~~}
		\multicolumn{4}{c}{\multirow{2}{*}{\textbf{Output Classification}} }\\
		\hhline{~~~~}
		\multicolumn{4}{c}{} \\
		\hhline{~--~}
		\multicolumn{1}{c}{\textbf{(N, m)}} & \multicolumn{1}{|c|}{\textit{Ind.}}	& \multicolumn{1}{|c|}{\textit{Unif.}}	& \multicolumn{1}{c}{}\\
		\hhline{-==-}
		\multicolumn{1}{|c}{\multirow{2}{*}{\textbf{(3,13)}}} & \multicolumn{1}{|c|}{\textit{10}}	& \multicolumn{1}{|c|}{\textit{0}}	& \multicolumn{1}{c|}{\textit{Ind.}}\\
		\hhline{~---}
		\multicolumn{1}{|c}{\multirow{2}{*}{}} & \multicolumn{1}{|c|}{\textit{0}}	& \multicolumn{1}{|c|}{\textit{10}}	& \multicolumn{1}{c|}{\textit{Unif.}}\\
		\hhline{====}
		\multicolumn{1}{|c}{\multirow{2}{*}{\textbf{(5,50)}}} & \multicolumn{1}{|c|}{\textit{10}}	& \multicolumn{1}{|c|}{\textit{0}}	& \multicolumn{1}{c|}{\textit{Ind.}}\\
		\hhline{~---}
		\multicolumn{1}{|c}{\multirow{2}{*}{}} & \multicolumn{1}{|c|}{\textit{0}}	& \multicolumn{1}{|c|}{\textit{10}}& \multicolumn{1}{c|}{\textit{Unif.}}\\
		\hhline{====}
		\multicolumn{1}{|c}{\multirow{2}{*}{\textbf{(6,50)}}} & \multicolumn{1}{|c|}{\textit{10}}	& \multicolumn{1}{|c|}{\textit{0}}		& \multicolumn{1}{c|}{\textit{Ind.}}\\
		\hhline{~---}
		\multicolumn{1}{|c}{\multirow{2}{*}{}} & \multicolumn{1}{|c|}{\textit{0}}	& \multicolumn{1}{|c|}{\textit{10}}	& \multicolumn{1}{c|}{\textit{Unif.}}\\
		\hhline{----}
\end{tabular}
\caption{{\bf Boson Sampling vs Mean-Field Sampling and Uniform Sampling}. Left Table: Confusion matrix showing the results obtained by applying single trial \textit{K-means clustering} initialized with \textit{K-means++} to distinguish between a \textit{bona fide} Boson Sampler and a Mean-Field Sampler. Right Table: Confusion matrix showing the results obtained by applying single trial \textit{K-means clustering} initialized with \textit{K-means++} to distinguish between a \textit{bona fide} Boson Sampler and a Uniform Sampler. In both cases the parameters that were used are the same that were tuned for the case $N=3$ and $m=13$, which gave the results shown in Tab.III of the Main text.}
\end{table}

\section*{SUPPLEMENTARY NOTE 5: EXPERIMENTAL APPARATUS}

The laser source of the experiment generates a pulsed 785 nm field, which is frequency doubled by exploiting a second-harmonic generation (SHG) process in a BiBO (beta bismute borate) crystal, generating a 392.5 nm pump beam. In the Boson Sampling experiment [S4], the pump is injected in a 2-mm thick BBO (beta barium borate) crystal cut for type-II phase matching. Four photons are produced by a second-order process, are spectrally selected by a 3-nm interference filter and are spatially separated according to their polarization state by a polarizing beam-splitter. One of the four photons is directly detected with an avalanche photodiode (detector $D_{\mathrm{T}}$ in Fig. 3 of the main text) and acts as a trigger for the experiment. The other three-photons are coupled into single-mode fibers, are prepared in the same polarization state by fiber polarization controllers (PC) and propagate through independent delay lines that are employed to adjust their relative delay. Finally, they are injected in input modes (6,7,8) of a 13-mode integrated interferometers by means of a single-mode fiber array (FA), and are collected after the evolution by a multi-mode FA. The output modes are then measured with avalanche photodiodes. The output signals are elaborated by an electronic acquisition system to identify three-fold coincidences at the output of the device, conditioned to the detection of the trigger photon at $D_{\mathrm{T}}$.

In the scattershot Boson Sampling experiment [S5], the pump beam is injected into six independent down-conversion sources, physically obtained with three 2-mm long BBO crystals. In crystal $C_{1}$, the two output photons of one pair after separation by the PBS are directly injected into input modes 6 and 8 of the integrated interferometer. The third photon is obtained probabilistically from the other pair generated by crystal $C_{1}$ and from the 4 pairs generated by crystals $C_{2}$ and $C_{3}$. More specifically, the third photon is injected in a variable input mode identified by the detection of the twin photon in the corresponding trigger detector $D_{\mathrm{T}i}$. Furthermore, an optical switch is employed to increase the input state variability. The sampled input states are thus of the form (6,$j$,8), with $j=\{1,2,3,7,9,11,12,13\}$, for an overall set of 8 different input states.

\section*{SUPPLEMENTARY REFERENCES}

\noindent [S1]  J. MacQueen, Some methods for classification and analysis of multivariate observations. In of Calif. Press, U. (ed.) Proc.
Fifth Berkeley Symp. on Math. Statist. and Prob., vol. 1, 281 -
297 (1967).

\noindent [S2]  L. Bottou and Y. Bengio, Convergence properties of the k-means algorithms. In Advances in Neural Information Processing Systems, 7 (MIT press, 1995).

\noindent [S3]  M.C. Tichy, K. Mayer, A. Buchleitner, K. Molmer, Stringent and Efficient Assessment of Boson-Sampling Devices, Phys. Rev. Lett., vol 113, 020502 (2014).

\noindent [S4] N. Spagnolo, C. Vitelli, L. Sansoni, E. Maiorino, P. Mataloni, F. Sciarrino, D. J. Brod, E. F. Galvo, A. Crespi, R. Ramponi and R. Osellame, General rules for bosonic bunching in multimode
interferometers. Phys. Rev. Lett. 111, 130503 (2013).

\noindent [S5]  M. Bentivegna, N. Spagnolo,C. Vitelli,F. Flamini,N. Viggianiello, L. Latmiral,P. Mataloni,D. J. Brod,E. F. Galvo,R. Ramponi, R. Osellame and F. Sciarrino, Experimental scattershot
boson sampling. Science Advances 1, e1400255 (2015).